\RequirePackage{booktabs}
\documentclass[pdflatex,sn-mathphys-num]{sn-jnl}

\usepackage{graphicx}%
\usepackage{multirow}%
\usepackage{amsmath}
\usepackage{amsthm}%
\usepackage[title]{appendix}%
\usepackage{xcolor}%
\usepackage{textcomp}%
\usepackage{manyfoot}%
\usepackage{booktabs}%
\usepackage{algorithm}%
\usepackage{algorithmicx}%
\usepackage{algpseudocode}%
\usepackage{listings}%

\usepackage[bitstream-charter]{mathdesign}
\usepackage[utf8]{inputenc}
\usepackage{dsfont}
\usepackage{enumerate}
\usepackage{stmaryrd}
\usepackage[percent]{overpic}
\usepackage{apptools}
\usepackage{enumitem}

\AtAppendix{\counterwithin{theorem}{section}}

\newcommand{\ex}[1]{{\left\langle{#1}\right\rangle}}
\newcommand{\ket}[1]{\left|{#1}\right\rangle}
\newcommand{\bra}[1]{\left\langle{#1}\right|}
\newcommand{\brac}[1]{\left[\!#1\!\right]}
\newtheorem{theorem}{Assumption}
\newtheorem{objection}{Question}
\DeclareSymbolFont{usualmathcal}{OMS}{cmsy}{m}{n}
\DeclareSymbolFontAlphabet{\mathcal}{usualmathcal}

\theoremstyle{thmstyleone}%
\theoremstyle{thmstyletwo}%

\theoremstyle{thmstylethree}%

\begin{document}

\title[The Quantum Rashomon Effect]{An Extended Wigner's Friend Argument from Quantum Contextuality}


\author*[1]{\fnm{Jochen} \sur{Szangolies}}\email{jochen.szangolies@dlr.de}

\affil*[1]{\orgdiv{Institute for Software Technology}, \orgname{German Aerospace Center (DLR)}, \orgaddress{\street{Linder H\"{o}he}, \city{Cologne}, \postcode{51147}, \country{Germany}}}


\abstract{The Frauchiger-Renner argument aims to show that `quantum theory cannot consistently describe the use of itself': in many-party settings where agents are themselves subject to quantum experiments, agents may make predictions that contradict observations. Here, we introduce a simplified setting using only three agents, that is independent of the initial quantum state, thus eliminating in particular any need for entanglement, and furthermore does not need to invoke any final measurement and resulting collapse. Nevertheless, the predictions and observations made by the agents cannot be integrated into a single, consistent account. We propose that the existence of this sort of \emph{Rashomon effect}, i.e. the impossibility of uniting different perspectives, is due to failing to account for the limits put on the information available about any given system as encapsulated in the notion of an \emph{epistemic horizon}.}




\maketitle

\section{Introduction}

In Akira Kurosawa's 1950 film classic \emph{Rashomon}, the murder of a samurai and rape of his wife are recounted by different eyewitnesses, who give widely varying accounts of the events. Upon the viewer, it is thus impressed that the testimony of all, safe maybe one, must be false---there can be only one true account of events, and those who stray from it are either mistaken, or outright lying. Perhaps some are right about certain details, and wrong about others: the true story then would be a synthesis of those presented. On this basis, in a sociological or anthropological context, the \emph{Rashomon effect} refers to the existence of multiple, divergent accounts of the same situation \cite{anderson2016rashomon}.

This seems to be a reasonable expectation in a classical world: there is one and only one true narrative regarding how things happen, and different narratives are, if truthful, compatible, and may be integrated into a more complete narrative. We may imagine a `view from nowhere' \cite{nagel1989view}, a perspective which eliminates every subjective aspect, and leaves only the objective facts of the matter. 

\begin{figure}[ht] 
 \centering
 \begin{overpic}[width=0.5\columnwidth]{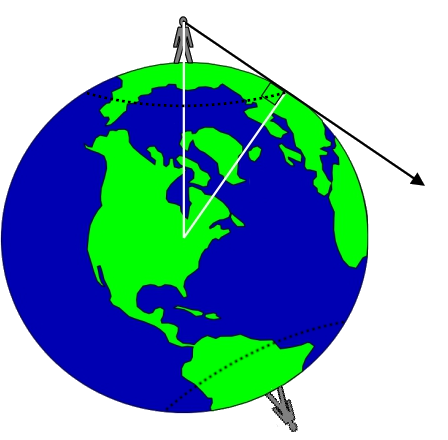}
  \put(35,92){$\mathcal{A}$}
  \put(68,4){$\mathcal{B}$}
\end{overpic}
\caption{Horizons as boundaries to simultaneously accessible information.}
\label{pic:horizons}
\end{figure}

But things may not always be this clear cut. Consider the horizon on the spherical Earth: there exists no point from which all Earth's surface is simultaneously in view (see Fig.~\ref{pic:horizons}). 

Two observers, $\mathcal{A}$ and $\mathcal{B}$, may have different horizons, different views of the same system (Earth or more accurately, its surface). Moreover, while we can imagine a `maximal' horizon by removing an observer's point of view infinitely far from the surface, this horizon would still not cover both the horizons of $\mathcal{A}$ and $\mathcal{B}$ completely.

Such a failure of the existence of a `global' perspective is nothing new in physics. In special relativity, there exists no global standard of simultaneity, and hence, no `global clock' keeping the true time. Indeed, the order of events in different frames of reference may differ: if two observers disagree about this order, this does not imply that one of them must be wrong.

Still, a failure of different perspectives to mesh can give us important insights regarding a theory's implications---both as regards the fundamental picture of the world it implies (its ontology), and in an operational sense, regarding what it permits, respectively forbids, any agent to do, or conclude. Therefore, the recent argument due to Frauchiger and Renner \cite{frauchiger2018quantum} is of great interest---as is evidenced by the large amount of research it has stimulated in a relatively short amount of time.

Frauchiger and Renner's titular claim is that `quantum mechanics cannot consistently describe the use of itself'---they argue, in brief, that different agents applying standard quantum theory to a special setting that itself includes further agents will derive contradictory conclusions: `one agent, upon observing a particular measurement outcome, must conclude that another agent has predicted the opposite outcome with certainty.' This calls the universal validity of quantum theory into question: there are situations in which the application of standard quantum theoretical methods apparently leads to false predictions.

We give a brief discussion of the Frauchiger-Renner argument, which is closely related to Hardy's paradox \cite{hardy1992quantum,hardy1993non}, in Sec.~\ref{wig}. We interpret this apparent inconsistency as due to a quantum version of the Rashomon effect: the existence of multiple narratives without the possibility of consistently combining them into a meta-narrative.

Then, in Sec.~\ref{qrash}, we propose a different extension of the `Wigner's Friend'-Ge\-dan\-ken\-ex\-peri\-ment: in a setting with three nested observers, inconsistent conclusions about observed experimental outcomes can be obtained \emph{without} any reference to a particular quantum state, and moreover, in a deterministic way. This shows that the quantum Rashomon effect does not depend on particular characteristics of a quantum system, such as the presence of entanglement, but is instead generic: for three observers, no `meta-narrative' exists integrating their individual viewpoints. The state-independence of this scenario, as well as the fact that one need not appeal to the projection postulate to obtain the contradiction, leads to the simultaneous presence of definite, yet irreconcilable narratives---irreconcilable `realities', in a certain sense.

Afterwards, we will analyze these arguments from the perspective of \emph{epistemic horizons} \cite{szangolies2018epistemic} in Sec.~\ref{epihor}. Such horizons pose limitations on the amount of information simultaneously available in a consistent way. They arise due to restrictions of paradoxical self-reference: as shown in Ref.~\cite{szangolies2018epistemic}, the assumption that all possible experimental outcomes---and hence, full information about a system---are simultaneously available can be brought to a contradiction via an appeal to Lawvere's fixed-point theorem \cite{lawvere1969diagonal}.

The presence of epistemic horizons---and thus, a maximum amount of information available about any given system---can be seen to give rise to many characteristically quantum phenomena, such as the superposition principle, the unpredictability of measurement outcomes, state-change upon measurement, complementarity, and the uncertainty principle \cite{szangolies2018epistemic}. Moreover, the concept can also be used to shed new light on classic controversies, such as the question of whether quantum mechanics is complete: applied to the reasoning of Einstein, Podolski, and Rosen (EPR) \cite{einstein1935can}, the incompleteness of quantum mechanics can only be concluded by means of an inconsistent counterfactual inference \cite{szangolies2020epi}. 

We demonstrate how the notion of epistemic horizons can profitably be applied to the interpretation of the Frauchiger-Renner argument. To this end, a particular emphasis is placed on the notion of \emph{conditional properties}, which will be seen to play a key role in our discussion of the Frauchiger-Renner argument, similarly to the `relative facts' of Di Biagio and Rovelli \cite{di2020stable}, or Bruckner's `observer-dependent facts' \cite{brukner2018no}---but without, arguably, a reliance on the notoriously hard to pin down concept of `observer'.

Finally, we conclude in Sec.~\ref{conc}.

\section{Wigner's Many Friends} \label{wig}

We start our investigation by considering the argument due to Frauchiger and Renner \cite{frauchiger2018quantum}, which they propose shows that `quantum theory cannot consistently describe the use of itself'. 

The Frauchiger-Renner argument is most simply explained as a `Wigner's Friendification' of Hardy's paradox (for details, see \cite{dourdent2020quantum,szangolies2020epi}). The procedure of `Wigner's Friendification' can be thought of as a means to eliminate the necessity for counterfactual reasoning from quantum arguments: instead of reasoning about what \emph{would} have happened, \emph{had} a measurement in a different, incompatible basis been performed, different bases are assigned to different `levels' of observers. By thus stipulating that each observer has obtained a definite outcome, we can appeal to these incompatible values as simultaneously co-existing `facts of the world'---or `elements of reality', in the oft-quoted phrasing of EPR \cite{einstein1935can}. 

\begin{figure}[h] 
 \centering
 \begin{overpic}[width=0.6\textwidth]{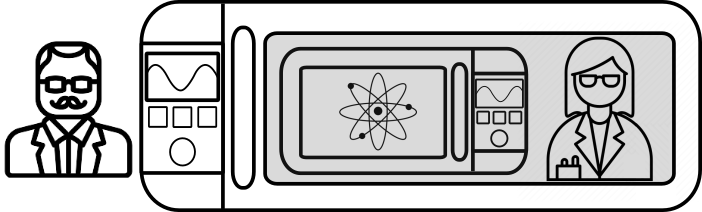}
  \put(15,21){$\mathcal{W}$}
  \put(91,21){$\mathcal{F}$}
\end{overpic}
\caption{Original `Wigner's Friend'-setup: The friend $\mathcal{F}$ performs a measurement on a quantum system, after which Wigner $\mathcal{W}$ performs a measurement on the total system containing $\mathcal{F}$ and the quantum system.}
\label{pic:WF}
\end{figure}

To see how this works, consider Wigner's original thought experiment \cite{wigner1995remarks}. Wigner $\mathcal{W}$ imagines an oberver $\mathcal{F}$, the eponymous `friend', carrying out a $z$-basis measurement on a system in the state
\begin{equation*}
    \ket{\Psi} = \frac{1}{\sqrt{2}}\left(\ket{z_\mathcal{F}^+}+\ket{z_\mathcal{F}^-}\right).
\end{equation*}

Here, the symbol $z_\mathcal{F}^\pm$ labels the $\pm$-Eigenstate of the $Z$-observable on the system under the direct control of the friend $\mathcal{F}$.

As no information leaks out of the lab to $\mathcal{W}$, he must describe the experiment by means of the unitary evolution
\begin{equation}\label{WF}
    \ket{\mathcal{F}}\otimes\frac{1}{\sqrt{2}}\left(\ket{z_\mathcal{F}^+}+\ket{z_\mathcal{F}^-}\right)\leadsto \frac{1}{\sqrt{2}}\left(\ket{f^+z_\mathcal{F}^+}+\ket{f^-z_\mathcal{F}^-}\right),
\end{equation}
where $f^\pm$ labels $\mathcal{F}$'s memory register.

We then define $\mathcal{W}$'s `$x$-basis' as:
\begin{equation}
    \left\{\ket{\brac{x}_\mathcal{W}^+},\ket{\brac{x}_\mathcal{W}^-}\right\} = \left\{\frac{1}{\sqrt{2}}\left(\ket{f^+z_\mathcal{F}^+}+\ket{f^-z_\mathcal{F}^-}\right),\frac{1}{\sqrt{2}}\left(\ket{f^+z_\mathcal{F}^+}-\ket{f^-z_\mathcal{F}^-}\right)\right\}
\end{equation}

Here, we have introduced the convention to indicate `Wigner's Friendification' by square brackets around a symbol, indicating the surrounding laboratory. We will also refer to quantities on this level as `boxed' observables. 

Friendification or boxing can then be understood as entangling an ancilla-qubit to a physical system in a `pre-measurement' step. Consider a lower-level observer $\mathcal{F}$ with access to a memory-qubit $f$ used to track the measurement outcome in the $z_\mathcal{F}$-basis:
\begin{equation}
    \ket{\mathcal{F}}\otimes\ket{z_\mathcal{F}^\pm}\leadsto \ket{f^\pm z_\mathcal{F}^\pm}\equiv \ket{\brac{z}_\mathcal{W}^\pm},
\end{equation}

where $\mathcal{W}$ refers to a higher-level (`Wigner') observer (outside the box).

Boxing is not limited to single observables. Measuring multiple (bivalent) observables on different qubits can be modeled by adding an equal number of register qubits, e.g.:

\begin{equation}
    \ket{\mathcal{F}}\otimes\ket{x_1^+z_2^-}\leadsto \ket{f_1^+f_2^-x_1^+z_2^-}\equiv \ket{\brac{x}_1^+\brac{z}_2^-},
\end{equation}

where the measurement basis is indicated by the associated register and boxing is performed per system/register qubit pair, e.g. $\ket{\brac{x}_1^+} = \ket{l_1^+x_1^+}$. Furthermore, boxing can be iterated over intermediary observers $\mathcal{M}$ by simply entangling further memory qubits, e.g.:

\begin{equation}
    \ket{\brac{\brac{x}}_{i,\mathcal{W}}^\pm}=\ket{m_i^\pm\brac{x}_{i,\mathcal{M}}}=\ket{m_i^\pm f_i^\pm x_{i,\mathcal{F}}^\pm},
\end{equation}
where $m_i$ denotes the $i$-th memory qubit of the intermediary observer $\mathcal{M}$.

It is obvious, but worth emphasizing that measurements in the boxed basis obey the same statistics as measurements in the original, unboxed basis on a given system. Take an arbitrary system in the $z$-basis, and let $\mathcal{F}$ carry out their measurement:
\begin{align}
    \nonumber\ket{\mathcal{F}}\otimes\ket{\Psi} & = \ket{\mathcal{F}}\otimes\left(\alpha^+\ket{z^+_\mathcal{F}} + \alpha^-\ket{z^-_\mathcal{F}}\right)\\
    & \label{prob}\leadsto \left(\alpha^+\ket{f^+z^+_\mathcal{F}} + \alpha^-\ket{f^-z^-_\mathcal{F}}\right) \\
    &\nonumber = \frac{\alpha^+ + \alpha^-}{\sqrt{2}}\ket{\brac{x}_\mathcal{W}^+} + \frac{\alpha^+ - \alpha^-}{\sqrt{2}}\ket{\brac{x}_\mathcal{W}^-},
\end{align}

yielding the same coefficients as writing the original $\ket{\Psi}$ in the (unboxed) $x_\mathcal{F}$-basis.

Consequently, carrying out a measurement in this basis will convince Wig\-ner that the laboratory, including the friend $\mathcal{F}$, must be in a superposed state; nevertheless, $\mathcal{F}$ may inform $\mathcal{W}$ that they have obtained \emph{some} definite outcome (albeit without revealing which outcome) \cite{deutsch1985quantum}. However, measurement in $\mathcal{W}$'s $\brac{x}_\mathcal{W}$-basis will leave the state on the right side of Eq.~\eqref{WF} invariant. Then, after this measurement, $\mathcal{W}$ can carry out a measurement in the $\brac{z}_\mathcal{W}$-basis $\{\ket{f^+}\otimes\ket{z_\mathcal{F}^+}, \ket{f^-}\otimes\ket{z_\mathcal{F}^-}\}$, by measuring

\begin{equation}
    \brac{Z}_\mathcal{W} = \ket{\mathcal{F}^+z_\mathcal{F}^+}\bra{\mathcal{F}^+z_\mathcal{F}^+} - \ket{\mathcal{F}^-z_\mathcal{F}^-}\bra{\mathcal{F}^-z_\mathcal{F}^-}.
\end{equation}

The outcome of this measurement corresponds to the truth value of the proposition `$\mathcal{F}$ has observed outcome $z_\mathcal{F}^+$' (for clarity, we may sometimes write $z_\mathcal{F}^+$ as $v(Z_\mathcal{F}) = +1$). As observed in \cite{healey2018quantum}, if measurement outcomes have any claim to objectivity, we should expect this to faithfully reveal $\mathcal{F}$'s observed value---that is, $\mathcal{W}$'s measurement of $\brac{Z}_\mathcal{W}$ reproduces $\mathcal{F}$'s measurement of $Z_\mathcal{F}$. Consequently, after $\mathcal{F}$ finds the system in, say, the state $\ket{z_\mathcal{F}^+}$, they should confidently predict that $\mathcal{W}$ will find $\ket{\brac{z}_\mathcal{W}^+}=\ket{f^+z_\mathcal{F}^+}$ upon performing the $\brac{Z}_\mathcal{W}$-measurement.

We take this as motivation for the following central assumption:

\begin{theorem}[Consistency of Observed Outcomes (COO)]
\label{assump:nonc}
For any two observables $O$ and $\brac{O}$, where $\brac{O}$ is the Wigner's Friendification of $O$, if $v(O) = o$ has been observed, then $v(\brac{O}) = o$ with certainty, independently of other compatible measurements performed on the system where $\brac{O}$ is defined. Furthermore, if $v(\brac{O}) = o$ implies $v(O^\prime) = o^\prime$, and $v(O) = o$ has been observed, then $v(O^\prime) = o^\prime$.
\end{theorem}

In other words, if $\mathcal{F}$ has observed $v(O) = o$, then they will be able to predict, with certainty, that the outcome of $\mathcal{W}$'s measurement of $\brac{O}$ will likewise be $o$. Appealing to the famous EPR criterion of reality, which proclaims that \cite{einstein1935can}
\begin{quote}
[i]f, without in any way disturbing a system, we can predict with certainty (i.e., with probability equal to unity) the value of a physical quantity, then there exists an element of reality corresponding to that quantity,
\end{quote}
$\mathcal{F}$ is then justified in positing the existence of an element of reality corresponding to $\brac{O}$.

This is then the second great boon of Friendification: it allows to view a measurement, having been carried out behind closed doors, so to speak, to be modeled by a unitary evolution from the outside observer's point of view, and thus, the simultaneous measurement of `complementary' observables.

At the same time that, for Wigner, there is a fact of the matter that the laboratory is in a state described by $\brac{x}_\mathcal{W}^+$, for $\mathcal{F}$, there must be a definite fact regarding whether the system is described by $z_\mathcal{F}^+$ or $z_\mathcal{F}^-$. 

\begin{figure}[h] 
 \centering
 \begin{overpic}[width=0.7\textwidth]{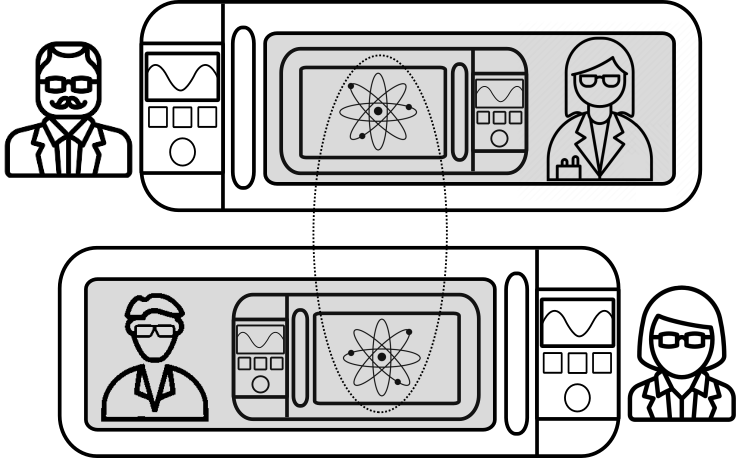}
  \put(14,52){$\mathcal{A}$}
  \put(85,52){$\mathcal{F_A}$}
  \put(97,19){$\mathcal{B}$}
  \put(25,19){$\mathcal{F_B}$}
  \put(48,30){$\ket{\Psi_H}$}
\end{overpic}
\caption{Frauchiger and Renner's elaboration of the Wigner's Friend-thought experiment: two `Wigners', $\mathcal{A}$ and $\mathcal{B}$, carry out measurements on two (entangled) labs containing their respective friends, $\mathcal{F_A}$ and $\mathcal{F_B}$.}
\label{pic:FR}
\end{figure}
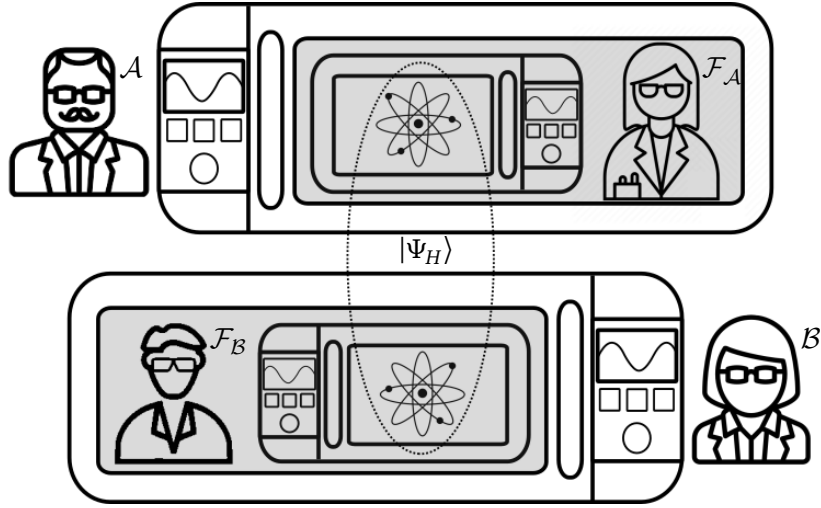

Frauchiger and Renner essentially apply the Wigner's Friend-scenario to the Hardy paradox. Consider then two friends, $\mathcal{F_A}$  and $\mathcal{F_B}$, sharing an entangled system, and their respective `Wigners', $\mathcal{A}$ and $\mathcal{B}$, as in Fig.~\ref{pic:FR}. The system shared by $\mathcal{F_A}$  and $\mathcal{F_B}$ is in the Hardy state

\begin{equation}\label{zz}
    \ket{\Psi_H}=\frac{1}{\sqrt{3}}\left(\ket{z_\mathcal{A}^+z_\mathcal{B}^+} + \ket{z_\mathcal{A}^+z_\mathcal{B}^-} + \ket{z_\mathcal{A}^-z_\mathcal{B}^+} \right).
\end{equation}

To simplify notation, we consider here an expression like `$\ket{z_\mathcal{A}^+}$' to be shorthand for `$\ket{f_A^+}\otimes\allowbreak\ket{z_\mathcal{A}^+}$', that is, a term in which $\mathcal{A}$'s friend $\mathcal{F_A}$ has carried out their $z$-measurement, and obtained the outcome $+1$ (\emph{mutatis mutandis} for the other expressions). 

The two observers $\mathcal{A}$ and $\mathcal{B}$ then carry out their measurements, as before, in the basis
\begin{equation}
\left\{\ket{\brac{x}^+},\ket{\brac{x}^-}\right\}=\left\{\frac{1}{\sqrt{2}}(\ket{z^+} + \ket{z^-}),\frac{1}{\sqrt{2}}(\ket{z^+} - \ket{z^-})\right\}.     
\end{equation}

Note that $\mathcal{A}$, $\mathcal{B}$, and their respective friends carry out the same measurement in each round of the experiment. Therefore, in contrast to Bell-like no-go theorems (e.g. \cite{brukner2018no,bong2020strong}), no `free will'-assumption is necessary, and the conclusion cannot be avoided by an appeal to superdeterminism.

Written in this basis, the combined state of the laboratories then is
\begin{equation}\label{xx}
    \ket{\Psi_H}= \frac{3}{\sqrt{12}}\ket{\brac{x}_\mathcal{A}^+\brac{x}_\mathcal{B}^+} +  \frac{1}{\sqrt{12}}\ket{\brac{x}_\mathcal{A}^+\brac{x}_\mathcal{B}^-} + 
     \frac{1}{\sqrt{12}}\ket{\brac{x}_\mathcal{A}^-\brac{x}_\mathcal{B}^+} - \frac{1}{\sqrt{12}}\ket{\brac{x}_\mathcal{A}^-\brac{x}_\mathcal{B}^-}.
\end{equation}
From this, the Born rule immediately tells us that both $\mathcal{A}$ and $\mathcal{B}$ may obtain the $-1$ outcome with probability $\frac{1}{12}$.

In order now to present the argument of Frauchiger and Renner, we introduce an additional assumption which allows us to simultaneously talk about the observations performed, and inferences made, by different observers. 

\begin{theorem}[Absoluteness of Reality (AR)]
\label{assump:abs}
For any two observers $\mathcal{A}$ and $\mathcal{B}$, if $\mathcal{A}$ is in a position to be certain that $\mathcal{B}$ can conclude the existence of an element of reality for an observable $O$, then $\mathcal{A}$ can likewise conclude the existence of an element of reality for $O$.
\end{theorem}

In other words, what's real for you is real for me. 

The two assumptions \textbf{COO} and \textbf{AR} are intended to formalize the intuition that there should be a single, unified account---a `reality' in the sense that refers to an assignment of definite values to observables---of all possible observations within the framework of quantum mechanics shared by all observers. A violation of these assumptions then indicates a lack of such an account, and hence, constitutes a Rashomon-like effect.

It should be noted here that \textbf{COO} is not theory-independent in its evaluation. The reason for this is that a determination which outcomes are implied by a given observed outcome is only possible on the basis of the predictions made by a given theory. In classical mechanics, for instance, there exist observables such that their values imply all outcomes of further measurements on a system, while no such determination is possible in the quantum world. Thus, different theories allow for different logical structures regarding the possible relations between outcomes.

Consequently, we can only ever evaluate \textbf{COO} relatively to a given theory. Any argument attempting to show a violation of the above assumptions then must appeal to the specifics of a given theory to show that, \textit{within that theory}, they cannot hold. This is fully analogous to cases like Bell's theorem \cite{bell1964einstein}, where it is the conjunction of `local realism' and the predictions of quantum theory that is inconsistent---and thus, the experimental confirmation of the quantum prediction entails a rejection of that assumption.

Frauchiger and Renner~\cite{frauchiger2018quantum} frame their discussion somewhat differently. Their argument starts with the following set of assumptions:
\begin{itemize}
    \item[(\textbf{Q})] Quantum theory is universally valid
    \item[(\textbf{C})] Agents' predictions should not be contradictory
    \item[(\textbf{S})] Each measurement has a single outcome from the point of view of the agent carrying out that measurement
\end{itemize}

Hence, the validity of quantum mechanics is an explicit assumption for their argument, which then entails that there is an inconsistency between their three assumptions. This is, however, largely a difference of emphasis: Frauchiger and Renner intend for their argument to show an inconsistency in the application of quantum mechanics to itself; the present aim here is to argue that, \textit{within} quantum mechanics, both \textbf{COO} and \textbf{AR} cannot hold. Consequently, my argument will appeal to the predictions of quantum mechanics to arrive at its conclusion.

There are certain relations between these assumptions and the ones used in the present argument. For instance, \textbf{Q} is, as noted, implied by the need to evaluate \textbf{COO} against theoretical predictions, such as to allow us to derive conclusions about observable values (and likewise, we only need those outcomes predicted to occur with probability 1). Similarly, \textbf{AR} plays a similar role to that of \textbf{C} in the original argument in that it allows one observer to promote another observer's conclusions to their own. 

Now we proceed to briefly outline the argument. In this, we will follow the presentation as given in Ref.~\cite{dourdent2020quantum}. The argument is recapitulated here for illustration only. For details, we refer to the original publication.
\begin{enumerate}[label=\roman*]
    \item\label{1} In the $\brac{x}_\mathcal{A}z_\mathcal{B}$-basis, the state is 
    \begin{equation}\label{xz}
        \ket{\Psi_H} = \sqrt{\frac{2}{3}}\ket{\brac{x}_\mathcal{A}^+z_\mathcal{B}^+} + \frac{1}{\sqrt{6}}\ket{\brac{x}_\mathcal{A}^+z_\mathcal{B}^-} + \frac{1}{\sqrt{6}}\ket{\brac{x}_\mathcal{A}^-z_\mathcal{B}^-}.
    \end{equation}
    Thus, if $\mathcal{A}$ obtains $-1$, from the $\ket{\brac{x}_\mathcal{A}^-z_\mathcal{B}^-}$-term, it follows that $\mathcal{F_B}$ must obtain $-1$, as well.
    \item\label{2} From \eqref{zz}, we then see that if $\mathcal{F_B}$ obtains $-1$, $\mathcal{F_A}$ must obtain the $+1$-outcome.
    \item\label{3} In the $z_\mathcal{A}\brac{x}_\mathcal{B}$-basis, the state is 
    \begin{equation}\label{zx}
        \ket{\Psi_H} = \sqrt{\frac{2}{3}}\ket{z_\mathcal{A}^+\brac{x}_\mathcal{B}^+} + \frac{1}{\sqrt{6}}\ket{z_\mathcal{A}^-\brac{x}_\mathcal{B}^+} + \frac{1}{\sqrt{6}}\ket{z_\mathcal{A}^-\brac{x}_\mathcal{B}^-}.
    \end{equation}
    Thus, we get that if $\mathcal{F_A}$ obtains the $+1$-outcome, $\mathcal{B}$ must likewise obtain the $+1$-outcome (from the $\ket{z_\mathcal{A}^+\brac{x}_\mathcal{B}^+}$-term).
    \item Putting \ref{1} - \ref{3} together, we thus obtain that if $\mathcal{A}$ obtains the $-1$-outcome, $\mathcal{B}$ obtains the $+1$ outcome.
    \item Yet, both observe the outcome $-1$ with probability $\frac{1}{12}$. $\lightning$
\end{enumerate}

Consequently, if $\mathcal{A}$ obtains the value $-1$ in their measurement, they know that $\mathcal{F_B}$ knows that $\mathcal{F_A}$ knows that $\mathcal{B}$ must obtain the outcome $+1$, and hence, they know themselves that $\mathcal{B}$ obtains that outcome. Yet, with probability $\frac{1}{12}$, both $\mathcal{A}$ and $\mathcal{B}$ observe the outcome $-1$.

The above derivation makes use of the original assumptions \textbf{Q}, \textbf{C} and \textbf{S}. However, it is instructive to see how these relate to the framework as presented here. Therefore note that the above can be seen to appeal to \textbf{COO} (Assumption~\ref{assump:nonc}) in several places: in step~\ref{1}, $\mathcal{A}$ takes their observation of the $-1$-value to determine that $\mathcal{F_B}$ must likewise obtain the $-1$-value. This is justified by an appeal to the quantum prediction: the only component of $\ket{\Psi_H}$ consistent with this observation is $\ket{\brac{x}_\mathcal{A}^-z_\mathcal{B}^-}$, which mandates a value of $-1$ for $\mathcal{F_B}$'s observation. In \ref{2} and \ref{3}, similar inferences are performed by $\mathcal{F_B}$ and $\mathcal{F_A}$. 

One should take care to note here that this reasoning does not include any appeal to the collapse dynamics (cf. the discussion in \cite{bub2020understanding}): it would be equally valid in, e.g., a relative state formulation of quantum mechanics. All that is necessitated is the consistency of further observations with those made by a given agent. In fact, if appeal were made to the collapse dynamics at this point, the argument would not go through, just as, in the original Wigner's Friend-scenario, appealing to the state collapse after the Friend's observation precludes a later detection of interference on the part of Wigner. In the following, formulations such as `$\mathcal{A}$ observes $-1$ for their measurement and concludes the state to be $\ket{\brac{x}_\mathcal{A}^-z_\mathcal{B}^-}$' should consequently be considered merely as shorthand for an invocation of \textbf{COO}, concluding that the only consistent outcome for a measurement of $z_\mathcal{B}$ given their observed value is $-1$.

Thanks to the Friendification, we are not talking about counterfactual observations here, but of actually obtained values of different observers. This, together with \textbf{AR} (Assumption~\ref{assump:abs}), allows the `daisy-chaining' of individual observer's conclusions into a single determination of $\mathcal{B}$'s value on the part of $\mathcal{A}$: to conclude, say, that $\mathcal{F_A}$ gets the outcome $+1$, $\mathcal{A}$ must posit that there is an element of reality assigned to $\mathcal{F_B}$'s measurement.

\section{A State-Independent Quantum Rashomon Effect}\label{qrash}

The Frauchiger-Renner argument confronts us with a surprising claim: quantum mechanical predictions, if extended to observers themselves using quantum theory, may yield contradictory situations. However, in putting the point forcefully, it suffers from certain drawbacks: it is probabilistic in nature, such that there is an element of chance to the emergence of conflict between different perspectives; and moreover, it relies on the presence of a quantum phenomenon we typically consider to be far removed from everyday experience, namely, quantum entanglement. 

Furthermore, to obtain the contradiction, in the end, the state must be projected onto the $\ket{\brac{x}_\mathcal{A}^-\brac{x}_\mathcal{B}^-}$-component. But this is no longer a state in which either of the friends is in an eigenstate of having made a definite observation; thus, the friends' knowledge of their respective systems is `undone' by the measurement performed on the system. This is, of course, a necessity to enforce ultimate consistency of the experimenters' observations, and does not detract from the force of conclusions formed before the final projection took place, but it does allow us to think of the experiment as appealing to one unified reality at any one given time---even if that reality is later `rolled back' by further experiments, due to the non-unitary change of state occurring at the final measurement.

The first drawback has already been addressed in \cite{brukner2018no}, making use of an argument based on a generalization of the Greenberger-Horne-Zeilinger argument against local realism \cite{greenberger1989going}. The argument there introduces a three-party `Wigner's Friend'-type scenario, three `Wigners' and three friends performing experiments on a threepartite entangled state. (Although the argument has been criticized as failing to establish its desired conclusion on the basis that on no repetition of the experiment, all observables are actually measured \cite{healey2018quantum}.)

In contrast, to address the further issues, I propose an argument that adds just a single additional `level' to Wigner's original scenario. The argument is based on the proof of the Kochen-Specker theorem \cite{kochen1975problem} due to Peres \cite{peres1990incompatible} and Mermin \cite{mermin1990simple}.

\begin{figure}[h] 
 \centering
 \begin{overpic}[width=0.9\textwidth]{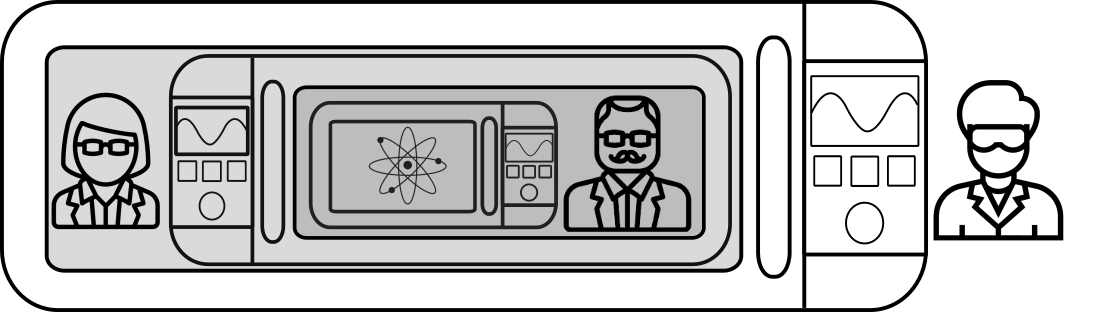}
  \put(61,18){$\mathcal{A}$}
  \put(13,18){$\mathcal{B}$}
  \put(96,18){$\mathcal{C}$}
\end{overpic}
\caption{The three-level `Wigner's Friend'-scenario: each of $\mathcal{A}$, $\mathcal{B}$, and $\mathcal{C}$ carries out three measurements according to Table~\ref{table:PM}.}
\label{pic:PM}
\end{figure}

The central object of the proof is the Peres-Mermin square of observables, as shown in Table~\ref{table:PM}.

\begin{table}[h]
\centering
\begin{tabular}{ |c|c|c| } 
 \cline{1-3}
  & &\\
 $\mathcal{A}_1 = Z_1\otimes\mathds{1}_2$ & $\mathcal{B}_1 = \mathds{1}_1\otimes \brac{Z}_2$ & $\mathcal{C}_1 = \brac{\brac{Z}}_1\otimes \brac{\brac{Z}}_2$\\
 & &\\
 \cline{1-3}
 & &\\
 $\mathcal{A}_2 = \mathds{1}_1\otimes X_2$ & $\mathcal{B}_2 = \brac{X}_1\otimes\mathds{1}_2$ & $\mathcal{C}_2 = \brac{\brac{X}}_1\otimes \brac{\brac{X}}_2$\\ 
 & &\\
 \cline{1-3}
 & &\\
 $\mathcal{A}_3 = Z_1\otimes X_2$ & $\mathcal{B}_3 = \brac{X}_1\otimes \brac{Z}_2$ & $\mathcal{C}_3 = \brac{\brac{Y}}_1\otimes \brac{\brac{Y}}_2$\\
 & &\\
 \cline{1-3} 
\end{tabular}
\caption{The Peres-Mermin square of observables, Wigner's-Friendified version.}
\label{table:PM}
\end{table}

For the \emph{unboxed} observables, implemented on a quantum system of dimension 4 in an arbitrary state, one readily checks that 

\begin{enumerate}
    \item\label{joint} Observables in each row and in each column commute,
    \item\label{constr} $\ex{\mathcal{A}_1\mathcal{B}_1\mathcal{C}_1} = \ex{\mathcal{A}_2\mathcal{B}_2\mathcal{C}_2} = \ex{\mathcal{A}_3\mathcal{B}_3\mathcal{C}_3} = 1$, and
    \item\label{constr2}$\ex{\mathcal{A}_1\mathcal{A}_2\mathcal{A}_3} = \ex{\mathcal{B}_1\mathcal{B}_2\mathcal{B}_3} = -\ex{\mathcal{C}_1\mathcal{C}_2\mathcal{C}_3} = 1$, 
\end{enumerate}
where $\ex{\ldots}$ denotes the expectation value.

The Kochen-Specker theorem can now be proven by noting that there exists no assignment of $\pm1$-values to the observables of the square that simultaneously satisfies the row- and column-constraints.

Now consider an arrangement of observers as in Fig.~\ref{pic:PM}. Here, we consider $\mathcal{A}$ to be the `innermost' observer, having direct access to a quantum system $\mathcal{S}$ in some arbitrary state $\ket{\Psi_\mathcal{S}}$, interrogating it by means of the measurements $\mathcal{A}_1$, $\mathcal{A}_2$, and $\mathcal{A}_3$. For simplicity, the system may be thought of as bipartite, with the indices on the measurements referring to the parts. $X$ is assumed to be the observable having $\ket{x^+}$ and $\ket{x^-}$ as eigenstates, and likewise for $Z$ and $Y$.

$\mathcal{B}$ is then the observer at the first level of `Wigner's Friendification'. Their observables refer, as before, to states of the system composed of the lab containing $\mathcal{A}$ and the object system. Likewise, $\mathcal{C}$ performs measurements on the lab containing $\mathcal{B}$ of appropriately constructed observables $\mathcal{C}_i$ (for the explicit construction of these observables, see Sec.~\ref{subsec:obj}).

\subsection{The Peres-Mermin-Wigner Argument}\label{subseq:PMW}

We will now present the main argument, and defer the discussion of possible open questions and objections to the sequel. The purpose of the argument is for $\mathcal{C}$ to derive information about $\mathcal{A}$'s outcomes, given their own outcomes as well as the row- and column-constraints of the Peres-Mermin square. As will be discussed, the row- and column-constraints implement the assumptions \textbf{COO} and \textbf{AR}, respectively.

The experiment proceeds according to the following steps:
\begin{enumerate}
    \item First, $\mathcal{A}$ performs their measurements, and signals completion to $\mathcal{B}$ (which can be done without disturbance, see above)
    \item Then, $\mathcal{B}$ performs their measurements, signalling completion to $\mathcal{C}$
    \item Finally, $\mathcal{C}$ performs their measurements
    \item $\mathcal{C}$ uses their outcomes to derive the number of $-1$-outcomes $\mathcal{A}$ must have obtained:
    \begin{enumerate}[label=\theenumi.\arabic*]
        \item Case 1: $\mathcal{C}$ has obtained $v(\mathcal{C}_i) = -1$ for all $i$:
        \begin{itemize}
            \item \textbf{AR}: There exist definite values (elements of reality) for all $\mathcal{B}_i$ such that
            \begin{equation*}
                \ex{\mathcal{B}_1\mathcal{B}_2\mathcal{B}_3} = v(\mathcal{B}_1)\cdot v(\mathcal{B}_2)\cdot v(\mathcal{B}_3) = 1. 
            \end{equation*}
            
            Consequently, $\mathcal{B}$ must have obtained zero or two values of $-1$.
            \item \textbf{COO}: $v(\mathcal{C}_i) = -1$ implies $v(\mathcal{A}_i)\neq v(\mathcal{B}_i)$, since
            \begin{equation*}
                \ex{\mathcal{A}_i\mathcal{B}_i\mathcal{C}_i} = v(\mathcal{A}_i)\cdot v(\mathcal{B}_i)\cdot v(\mathcal{C}_i) = 1.
            \end{equation*}
            
            \item Hence, $\mathcal{A}$ must obtain an odd number of $-1$-outcomes. However, by \textbf{AR}, there exist values for $\mathcal{A}_i$ such that $v(\mathcal{A}_1)\cdot v(\mathcal{A}_2)\cdot v(\mathcal{A}_3) = 1$, which necessitates an even number of $-1$-outcomes. $\lightning$
        \end{itemize}
        \item Case 2: $\mathcal{C}$ has obtained $v(\mathcal{C}_i) = -1$ for one $i$:
        \begin{itemize}
            \item \textbf{AR}: As above, $\mathcal{B}$ must have obtained either two or zero values of $-1$.
            \item \textbf{COO}: Again as before, $v(\mathcal{C}_i) = -1$ implies $v(\mathcal{B}_i)\neq v(\mathcal{A}_i$, while $v(\mathcal{C}_i) = -1$ implies $v(\mathcal{B}_i) = v(\mathcal{A}_i)$. Consequently, two of $\mathcal{A}$'s and $\mathcal{B}$'s values must agree, while one must differ.
            \item Suppose $\mathcal{B}$ has obtained zero $-1$-values. Consequently, $\mathcal{A}$ must have one. $\lightning$
            \item Then, let $\mathcal{B}$ have obtained $-1$ twice. If these are the rows where $\mathcal{C}_i = 1$, $\mathcal{A}$ must have obtained $-1$ three times, twice where it matches with the values of $\mathcal{B}$, and once where $\mathcal{B}$ has obtained $1$, and $\mathcal{C}$ has obtained $-1$. $\lightning$
            \item Alternatively, only one of $\mathcal{B}$'s two $-1$-values matches $\mathcal{A}$'s. Then, both of $\mathcal{A}$'s other values must be $1$, one being the same as $\mathcal{B}$'s, the other being opposite to $\mathcal{B}$'s second value of $-1$. This treats all cases, and in every case, $\mathcal{A}$ must obtain an odd number of $-1$-values. $\lightning$
        \end{itemize}
    \end{enumerate}
\end{enumerate}

This completes the basic argument: no matter what values are obtained by $\mathcal{C}$, together with the row- and column-constraints, $\mathcal{C}$ will predict an odd number of $-1$-outcomes for $\mathcal{A}$, yet $\mathcal{A}$ must obtain either zero or two. 

One way to avoid this result is to simply deny the applicability of quantum mechanics to systems that can be classified as `observers' in some appropriate sense. Indeed, this was Wigner's original conclusion, yielding the proposition that `consciousness causes the collapse'. If this is the case, then $\mathcal{B}$ cannot apply the unitary dynamics to the measurement process of $\mathcal{A}$, and consequently, the above argument fails to go through.

However, the present intent is to study what quantum mechanics, in its standard form, entails regarding the assumptions \textbf{COO} and \textbf{AR}. Therefore, we will postpone discussion of this possibility and some of its implications to Sec.~\ref{subsec:qcomp}.

\subsection{Elaborations on the Basic Argument}

The basic argument given in the preceding section can be elaborated on in several ways that serve to strengthen the result. For one, the proof appealed to the column constraint $\ex{\mathcal{B}_1\mathcal{B}_2\mathcal{B}_3} = 1$. However, we do not need to explicitly make use of this fact: the row constraints, alone, suffice to entail that only one out of $\mathcal{A}$ and $\mathcal{B}$ can have an odd number of $-1$-outcomes. For suppose all of $\mathcal{A}$, $\mathcal{B}$, and $\mathcal{C}$ had an odd number of $-1$-assignments: then, the product of all observed outcomes would be $-1$; however, due to the row-constraints, it must be $+1$. Hence, the argument can be made, instead, in reference to whoever has an even number of $-1$-outcomes, leading to the conclusion that one out of $\mathcal{A}$ and $\mathcal{B}$ must obtain an odd number of $-1$-outcomes, whereas in fact, neither do.

The preceding presentation had a `retrodictive' character: $\mathcal{C}$, having obtained their outcomes, reasons back to what $\mathcal{A}$'s outcomes must have been. In contrast, Frauchiger and Renner's original presentation emphasized a paradox of prediction: $\mathcal{B}$ may, in their example, conclude that $\mathcal{A}$ has predicted that they cannot observe $-1$, and thus, adopt the same conclusion---leading to a contradiction. The same can be done in this case: $\mathcal{C}$ may, before performing any measurement, conclude that $\mathcal{A}$ would predict them to observe an even number of $-1$-values, and then obtain a `contradiction' upon, in fact, obtaining an odd number.

To see this, first consider that $\mathcal{A}$ may obtain either zero or two $-1$-values. Consequently, they must predict that $\mathcal{B}$'s and $\mathcal{C}$'s values either agree in every row, or disagree in precisely two rows. If $\mathcal{B}$ now obtains an even number of $-1$-values, in the first case, it is immediately clear that so must $\mathcal{C}$. In the second case, if $\mathcal{B}$ obtains no $-1$-values, $\mathcal{C}$ must obtain two; while if $\mathcal{B}$ obtains two $-1$-values, $\mathcal{C}$ must obtain either none, or two, as well. Hence, $\mathcal{C}$ knows that $\mathcal{A}$ knows that $\mathcal{C}$ must observe an even number of $-1$-values, but will, in fact, observe an odd number.

So far, this essentially recapitulates the Frauchiger-Renner argument, albeit in a manner that does not depend on the initial quantum state, and does not involve a probabilistic component (and needs fewer observers). This does throw the issue in somewhat sharper psychological relief: by getting rid of the `spooky' influence of entanglement (between systems in distinct laboratories), the situation becomes that of a closed room mystery---in principle, if you are performing experiments within your laboratory, there could be observers right outside in the hall (and in front of the university building, say), whose experiences cannot be integrated with yours (provided all doors are sufficiently tightly shut): one could ask a set of questions of each, such that no consistent set of answers is possible. Moreover, while in the original thought experiment, a contradiction only obtains with probability 1 in the limit of many runs, in this version, each and every time the experiment is run, a contradiction will be derived.

However, the most surprising consequences of the above argument require some further teasing out. First, observe, as pointed out before, that each of $\mathcal{A}$ and $\mathcal{B}$ can communicate the fact of having made \emph{some} definite observation in accordance with their expectations to the next higher `rung' of the experiment, as long as they avoid mention of the details of their outcomes. Additionally, $\mathcal{C}$ will derive their contradiction in every term of the final superposed state. 

Suppose now that $\mathcal{C}$ has both a register for their obtained experimental outcomes $\ket{\mathcal{C}_1,\mathcal{C}_2,\mathcal{C}_3}$, and a register $\ket{\mathcal{M_C}}$ into which they note whether they, $\mathcal{A}$, and $\mathcal{B}$ have made some definite observation (which fact was communicated upwards through the levels of observers), and whether they have deduced a contradiction. The overall evolution of $\mathcal{C}$ then will be (where `$\ket{!_\mathcal{X}}$' denotes `$\mathcal{X}$ has made some definite observation', and `$\ket{\lightning_\mathcal{C}}$' denotes `$\mathcal{C}$ has derived a contradiction'):

\begin{align*}
    \ket{\mathcal{M_C}}\otimes&\ket{\mathcal{C}_1,\mathcal{C}_2,\mathcal{C}_3} \leadsto \\
        & \alpha\ket{!_\mathcal{A},!_\mathcal{B},!_\mathcal{C},\lightning_\mathcal{C}}\otimes\ket{\mathcal{C}_1^+,\mathcal{C}_2^+,\mathcal{C}_3^-} + 
         \beta\ket{!_\mathcal{A},!_\mathcal{B},!_\mathcal{C},\lightning_\mathcal{C}}\otimes\ket{\mathcal{C}_1^+,\mathcal{C}_2^-,\mathcal{C}_3^+} + \\
        & \gamma\ket{!_\mathcal{A},!_\mathcal{B},!_\mathcal{C},\lightning_\mathcal{C}}\otimes\ket{\mathcal{C}_1^-,\mathcal{C}_1^+,\mathcal{C}_3^+} + 
         \delta\ket{!_\mathcal{A},!_\mathcal{B},!_\mathcal{C},\lightning_\mathcal{C}}\otimes\ket{\mathcal{C}_1^-,\mathcal{C}_2^-,\mathcal{C}_3^-}, \\
\end{align*}

with normalization coefficients $\alpha$, $\beta$, $\gamma$ and $\delta$ depending on the original state. Consequently, the state of the memory register $\ket{\mathcal{M_C}}$ factorizes from the total state of $\mathcal{C}$. 

Now suppose we invite another friend, $\mathcal{D}$, to the party. They are again assumed to be outside of $\mathcal{C}$'s lab, which is sufficiently well shielded to maintain its coherence. The state of $\ket{\mathcal{M_C}}$ can now be communicated to $\mathcal{D}$, without influencing the superposition on the rest of the laboratory. Afterwards, $\mathcal{D}$, without making any measurement on the state of $\mathcal{C}$'s laboratory or its contents, then holds the following items of knowledge:
\begin{enumerate}
    \item $\mathcal{A}$ has signalled that they have made a definite observation
    \item $\mathcal{B}$ has signalled that they have made a definite observation
    \item $\mathcal{C}$ has signalled that they have made a definite observation
    \item $\mathcal{C}$ has signalled that, on the basis of their observations, they are certain that there is a contradiction between their observations and what $\mathcal{A}$, based on their observations, knows must be the case of $\mathcal{C}$'s observations
\end{enumerate}

Importantly, relatively to $\mathcal{D}$, no measurement has been performed on the state of the lab at this point; everything so far was merely unitary evolution. If $\mathcal{D}$ now assumes that $\mathcal{A}$, $\mathcal{B}$, and $\mathcal{C}$ signalling their having made a definite observation is sufficient grounds for believing that they have, in fact, made a definite observation, then $\mathcal{D}$ must conclude that each of them has obtained some definite set of values for their observations---yet, it is impossible to simultaneously assign values to observables in the experiment in such a way that they possibly all could have made definite observations (in accordance with their expectations). Furthermore, $\mathcal{D}$ knows that $\mathcal{C}$ is aware of the presence of this contradiction.

Thus, $\mathcal{D}$ would conclude that the sealed laboratory before them must contain mutually irreconcilable realities---one for $\mathcal{A}$'s definite values, one for $\mathcal{B}$'s definite values, and one for $\mathcal{C}$'s, but none for all of them at once. `Reality' is, of course, a difficult concept to pin down in this context---but for our purposes, it will suffice to use the definition implicit in the concept of `local realism' applied to Bell's theorem: that is, the simultaneous definiteness of observable values. In this sense, the Rashomon effect is a fundamental feature of quantum reality: there is no possible narrative that accounts for each of the observer's observations; yet, we have grounds to consider all of them equally real.

This is brought to its final consequence by supposing that there is, in fact, no $\mathcal{D}$---that indeed, $\mathcal{C}$ and their lab is the final level, and there is nothing beyond: that, in other words, the boundaries of $\mathcal{C}$'s lab constitute the edge of the universe. Then, there is nothing with respect to which the lab could decohere---and hence, we should imagine it as the state of the whole world, evolving unitarily. This world would then be one whose inhabitants, despite being able to communicate (within limits), cannot be said to wholly share the same reality (in the sense appealed to above).

The inhabitants of such a quantum universe, it seems, then may have definite, yet irreconcilable experiences---the world decomposes into a collection of perspectives, only ever partially overlapping, that resist integration into an all-encompassing `view from nowhere'. 

This is a conclusion that is stronger than either the original Frauchiger-Renner argument \cite{frauchiger2018quantum} or approaches based on Bell-like inequalities \cite{brukner2018no,healey2018quantum,bong2020strong}. The Frauchiger-Renner argument has to appeal to a final projection onto the $\ket{\brac{x}_\mathcal{A}^-\brac{x}_\mathcal{B}^-}$-component, while the above obtains even in a unitarily evolving quantum universe. The projection postulate is thus not an ingredient in the above argument, making it strictly stronger.

Bell-like approaches, by contrast, obtain their contradiction only in a statistical sense, across many repetitions of the same scenario---for \emph{any given} run of the experiment, no contradiction is obtained by all observer's values being objectively definite; only across an ensemble of runs can we conclude that not for \emph{every} run, such a distribution of objective values can exist. In the present argument, a contradiction is obtained in a single run.

Furthermore, as in the original Frauchiger-Renner argument, no assumption of `free choice' is needed, as in every run of the experiment, every observer measures all observables at their disposal. Frauchiger and Renner contrast their result with other approaches that do include such an assumption:
\begin{quote}
    Theorem 1 may be compared to earlier no-go results [...] which also use assumptions similar to (Q) and (S) (although the latter is often implicit). These two assumptions are usually shown to be in conflict with additional assumptions about reality, locality, or freedom of choice. [...] Here, we have shown that Assumptions (Q) and (S) are already problematic by themselves, in the sense that agents who use these assumptions to reason about each other [...] will arrive at inconsistent conclusions. [...] In fact, in the proposed experiment, the agents never make any choices.
\end{quote}

Thus, as opposed to the Bell-like approaches, an appeal to `superdeterminism' cannot avert the conclusion in this case.

\subsection{Quantum Computing and Experimental Metaphysics}\label{subsec:qcomp}

Bell's theorem has famously been referred to as `experimental metaphysics' by Abner Shimony \cite{shimony1989search}, bringing a question of ontology---whether the world is `locally realistic' in Bell's sense---into the laboratory. Far from a purely academic pursuit, it was this work that ultimately helped start the quantum revolution, leading to the development of secure techniques of quantum cryptography, quantum sensing, and providing important foundational impulses to the theory of quantum computation, eventually being honored with the 2022 physics Nobel prize.

In contrast, the possibility of experimental realization of extended Wigner's Friend-type experiments seems, at present, rather ludicrous, requiring the quantum superposition of entire laboratories. This opens up a question, however, of what sort of system qualifies as an `observer' in quantum mechanics. At one end of the scale, we as human beings certainly suffice. However, it seems implausible that we are at the minimum necessary complexity. This is a question already pondered by Bell \cite{bell2001quantum}: 

\begin{quote}
    What exactly qualifies some physical systems to play the role of `measurer'? Was the wavefunction of the world waiting to jump for thousands of millions of years until a single-celled living creature appeared? Or did it have to wait a little longer, for some better qualified system... with a PhD?
\end{quote}

At the extreme opposite end of the spectrum, one might hold that every interaction qualifies as observation, and thus, every physical system can play the role of an observer. Under such an assumption, extended Wigner's Friend-type experiments can be implemented, and indeed already have been \cite{proietti2019experimental, bong2020strong}. However, between these two extremes, there remains a huge middle ground left to explore. 

One motivation for such exploration may be derived from modifications of quantum theory that introduce a nonlinear collapse dynamics based on some measure of internal complexity \cite{kremnizer2015integrated, okon2020consciousness, chalmers2022consciousness}, e.g. (an appropriately modified version of) integrated information \cite{albantakis2023integrated}. Under such a proposal, an increase in the internal complexity of a prospective `observer' will lead to deviations from the predictions of ordinary quantum mechanics as the likelihood of objective wave function collapse reaches appreciable levels, thus providing a potential test of these models. 

A quantum computer forms a natural environment to carry out such tests. Indeed, shortly before submitting his seminal paper on the possibility of a quantum computer \cite{deutsch1985quantumc}, Deutsch was contemplating the puzzle posed by Wigner's thought experiment \cite{deutsch1985quantum}. Thus, should any deviations from ordinary quantum mechanics exist, there are potential implications on the feasibility of large-scale quantum computing: when the information integrated by a quantum program approaches the collapse threshold, the resulting loss of coherence might provide a complexity boundary to the implementation of certain algorithms. Thus, studying the feasibility of carrying out extended Wigner's Friend-type experiments within a quantum computer may serve as a tool to probe possible limits of quantum computation.

Along these lines, Wiseman et al. have introduced the notion of `thoughtfulness' \cite{wiseman2023thoughtful}, a set of assumptions requiring that `thought' supervenes on physical events, that communicable thoughts are absolutely real, and that any system outwardly displaying apparent cognitive abilities on par with a human observer should be considered to possess equivalent mental capacities. They then introduce a hypothetical human-level AI agent they christen \textsc{Quall-E}, intended to replace one of the three observers (at the level of a `friend') needed for their protocol and implemented on a sufficiently powerful quantum computer. 

The resulting experiment is based on the earlier `local friendliness'-theorem \cite{bong2020strong}. As such, it requires an assumption of locality (`local agency', requiring interventions at spacelike distances to be uncorrelated). Furthermore, an entangled state needs to be prepared and distributed over a sufficient distance. Then, the experiment needs to be repeated an appropriate number of times to gather statistics to evaluate the violation of a particular inequality. The implementation of something like the original Frauchiger-Renner setup would even require entangling two quantum computers with one another.

In contrast, a similar experiment based on the extended Wigner's Friend scenario exhibited here needs no assumption of locality, and a contradiction with its assumptions is obtained upon a single run. However, an added complication is that it would need at minimum two nested \textsc{Quall-E}-like agents, one of which carries out measurements on the other. Nevertheless, one might propose starting with simpler agents, increasing their degree of complexity as technology advances. A minimal requirement might be the implementation of a logic capable of reproducing the argumentation leading to the expectations on the measurement outcomes of other observers. However, we leave a detailed investigation of this proposal for future work.

\subsection{Some Possible Questions}\label{subsec:obj}

\begin{objection}[Doubly-Friendified Observables]
    \textbf{COO} connects values of observables $O$ with their Friendified versions $\brac{O}$. But $\mathcal{C}$ exists at a second level of Friendification. Are there appropriate observables such that $v(O) = v\left(\brac{\brac{O}}\right)$?
\end{objection}

Such observables can be constructed as follows. $\mathcal{A}$, after measurement, can be in one of four different states, from $\ket{\mathcal{A}_1^+,\mathcal{A}_2^+,\mathcal{A}_3^+}$ to $\ket{\mathcal{A}_1^-,\mathcal{A}_2^-,\mathcal{A}_3^+}$. We can thus index these states by a two-qubit system, consisting of qubits $a_1$ and $a_2$ ($\mathcal{A}$'s `memory'). Thus, $\mathcal{A}$'s possible post-measurement states are:
\begin{align*}
    \ket{\mathcal{A}_1^+,\mathcal{A}_2^+,\mathcal{A}_3^+} & = \ket{a_1^+a_2^+}\\
    \ket{\mathcal{A}_1^+,\mathcal{A}_2^-,\mathcal{A}_3^-} & = \ket{a_1^+a_2^-}\\
    \ket{\mathcal{A}_1^-,\mathcal{A}_2^+,\mathcal{A}_3^-} & = \ket{a_1^-a_2^+}\\
    \ket{\mathcal{A}_1^-,\mathcal{A}_2^-,\mathcal{A}_3^+} & = \ket{a_1^-a_2^-}
\end{align*}

Using this identification, we can write down the eigenstates of $\mathcal{B}$'s observables. For the $\brac{X}_1$ observable, we have:
\begin{align*}
    \ket{\brac{x}_1^+}&=\frac{1}{\sqrt{2}}\left(\ket{a_1^+z_1^+}+\ket{a_1^-z_1^-}\right)\\
    \ket{\brac{x}_1^-}&=\frac{1}{\sqrt{2}}\left(\ket{a_1^+z_1^+}-\ket{a_1^-z_1^-}\right),
\end{align*}
and analogously, for the $\brac{Z}_2$ observable:
\begin{align*}
    \ket{\brac{z}_2^+}&=\frac{1}{\sqrt{2}}\left(\ket{a_2^+x_2^+}+\ket{a_2^-x_2^-}\right)\\
    \ket{\brac{z}_2^-}&=\frac{1}{\sqrt{2}}\left(\ket{a_2^+x_2^+}-\ket{a_2^-x_2^-}\right).
\end{align*}

At the final level of `Wigner's Friendification' stands $\mathcal{C}$, performing measurements on the lab containing $\mathcal{B}$ and $\mathcal{A}$'s lab. Their observables can be constructed in the following way. For the $\brac{\brac{X}}_1$-observable, eigenstates are obtained in the by now familiar manner of Wigner's Friendification (where $b_1$ and $b_2$ again denote the `memory' qubits of $\mathcal{B}$):

\begin{align*}
    \ket{\brac{\brac{x}}_1^+} & = \ket{b_1^+\brac{x}_1^+}\\
    \ket{\brac{\brac{x}}_1^-} & = \ket{b_1^-\brac{x}_1^-}
\end{align*}
From this, we can then define the $\brac{\brac{Z}}_1$- and $\brac{\brac{Y}}_1$-eigenstates in the usual fashion:
\begin{align*}
    \ket{\brac{\brac{z}}_1^\pm} & = \frac{1}{\sqrt{2}}\left(\ket{\brac{\brac{x}}_1^+} \pm \ket{\brac{\brac{x}}_1^-} \right) \\
    \ket{\brac{\brac{y}}_1^\pm} & = \frac{1}{\sqrt{2}}\left(\ket{\brac{\brac{z}}_1^+} \pm i\ket{\brac{\brac{z}}_1^-} \right)
\end{align*}
For the observables $\brac{\brac{X}}_2$, $\brac{\brac{Y}}_2$, and $\brac{\brac{Z}}_2$, we proceed in the same fashion, starting with $\ket{\brac{\brac{z}}_2^\pm} = \ket{b_2^\pm\brac{z}_2^\pm}$.

One can readily check that for these observables, 
\begin{equation*}
    \brac{\brac{Z}}_j \cdot \brac{\brac{X}}_j = -i\brac{\brac{Y}}_j
\end{equation*}
for $j=1,2$. Furthermore, $\mathcal{C}_1$, $\mathcal{C}_2$, and $\mathcal{C}_3$ mutually commute. This implies that the column constraint $\ex{\mathcal{C}_1\mathcal{C}_2\mathcal{C}_3} = -1$ is satisfied for the boxed observables.

Likewise, it is important to note that the row-constraints in~\ref{constr} are obeyed for the boxed observables. In fact, this is how \textbf{COO} (Assumption~\ref{assump:nonc}) is implemented in the scenario.

To see this, suppose for the moment that $\mathcal{A}$ has observed the values $\mathcal{A}_1 = +1$, $\mathcal{A}_2 = -1$, and $\mathcal{A}_3 = -1$. Thus, they conclude that the system $\mathcal{S}$ is in the state
\begin{equation}
    \ket{\Psi_\mathcal{S}} = \ket{z_1^+x_2^-}.
\end{equation}

After $\mathcal{A}$'s measurement, the total system composed of $\mathcal{A}$ and $\mathcal{S}$ will then be in the state:
\begin{equation}
    \ket{\Psi_\mathcal{AS}} = \ket{a_1^+a_2^-z_1^+x_2^-}
\end{equation}

Written in $\mathcal{B}$'s measurement basis, this state is:
\begin{equation}\label{Bbase}
    \ket{\Psi_\mathcal{AS}} = 
    \frac{1}{2}\left(\ket{\brac{x}_1^+\brac{z}_2^+} - \ket{\brac{x}_1^+\brac{z}_2^-} + \ket{\brac{x}_1^-\brac{z}_2^+} - \ket{\brac{x}_1^-\brac{z}_2^-}\right).
\end{equation}

Now let $\mathcal{B}$ carry out their measurements. For each of the components in Eq.~\eqref{Bbase}, we obtain, in the $\brac{\brac{z}}_1\brac{\brac{x}}_2$-basis:
\begin{align}
    \label{x+z+}\ket{b_1^+b_2^+\brac{x}_1^+\brac{z}_2^+} = &
    \frac{1}{2}\left(\ket{\brac{\brac{z}}_1^+\brac{\brac{x}}_2^+} + \ket{\brac{\brac{z}}_1^+\brac{\brac{x}}_2^-}+\ket{\brac{\brac{z}}_1^-\brac{\brac{x}}_2^+} + \ket{\brac{\brac{z}}_1^-\brac{\brac{x}}_2^-}\right)\\
    \label{x+z-}\ket{b_1^+b_2^-\brac{x}_1^+\brac{z}_2^-} = &
    \frac{1}{2}\left(\ket{\brac{\brac{z}}_1^+\brac{\brac{x}}_2^+} - \ket{\brac{\brac{z}}_1^+\brac{\brac{x}}_2^-}+\ket{\brac{\brac{z}}_1^-\brac{\brac{x}}_2^+} - \ket{\brac{\brac{z}}_1^-\brac{\brac{x}}_2^-}\right)\\
    \label{x-z+}\ket{b_1^-b_2^+\brac{x}_1^-\brac{z}_2^+} = &
    \frac{1}{2}\left(\ket{\brac{\brac{z}}_1^+\brac{\brac{x}}_2^+} + \ket{\brac{\brac{z}}_1^+\brac{\brac{x}}_2^-} -\ket{\brac{\brac{z}}_1^-\brac{\brac{x}}_2^+} - \ket{\brac{\brac{z}}_1^-\brac{\brac{x}}_2^-}\right)\\
    \label{x-z-}\ket{b_1^-b_2^-\brac{x}_1^-\brac{z}_2^-} = &
    \frac{1}{2}\left(\ket{\brac{\brac{z}}_1^+\brac{\brac{x}}_2^+} - \ket{\brac{\brac{z}}_1^+\brac{\brac{x}}_2^-}-\ket{\brac{\brac{z}}_1^-\brac{\brac{x}}_2^+} + \ket{\brac{\brac{z}}_1^-\brac{\brac{x}}_2^-}\right)
\end{align}

Consequently, the state in that basis finally is

\begin{equation}\label{BAS}
    \ket{\Psi_\mathcal{BAS}} = \frac{1}{2}\left(\eqref{x+z+} - \eqref{x+z-} + \eqref{x-z+} - \eqref{x-z-}\right) = \ket{\brac{\brac{z}}_1^+\brac{\brac{x}}_2^-}.
\end{equation}

Hence, measuring $\brac{\brac{Z}}_1$ and $\brac{\brac{X}}_2$ recapitulates $\mathcal{A}$'s measurement, in this particular case. The same argument can then be carried through for the states $\ket{z_1^+x_2^+}$, $\ket{z_1^-x_2^+}$, and $\ket{z_1^-x_2^-}$, and thus, for all possible combinations of outcomes of $\mathcal{A}$. Consequently, if $\mathcal{C}$ were to carry out doubly-Friendified versions of $\mathcal{A}$'s measurements, they would recover $\mathcal{A}$'s outcomes, as stipulated by \textbf{COO}. $\mathcal{A}$ is hence justified in concluding the existence of elements of reality for $\mathcal{C}$'s observables.

\begin{objection}[Incompatibility of Friendified Observables]
    $\mathcal{A}$, upon measuring their observables, may be certain that doubly-Friendified versions recapitulate their values. However, $\mathcal{C}$ implements measurements $\mathcal{C}_i$ that are incompatible with $\brac{\brac{\mathcal{A}}}_j$ (and $\brac{\mathcal{B}}_j$) for $j\neq i$. Can one still conclude that the row-constraints hold, in this case?
\end{objection}

After $\mathcal{A}$ measures their observables $\mathcal{A}_i$, as we have seen, they are licensed to conclude the existence of elements of reality for the doubly-boxed observables $\mathcal{\brac{\brac{A}}}_i$ with values $v\left(\mathcal{\brac{\brac{A}}}_i\right) = v(\mathcal{A}_i)$. This is fundamentally the same reasoning as carried out in the Frauchiger-Renner argument in, e.g., step~\ref{1}: there, after $\mathcal{A}$ has made the measurement in the $\brac{x}$-basis and obtained the outcome $-1$, they conclude the state to be $\ket{\brac{x}_\mathcal{A}^-z_\mathcal{B}^-}$, thus concluding that there must exist an element of reality corresponding to $\mathcal{F_B}$'s $z$-measurement yielding the outcome $-1$.

This reasoning carries over directly from the original Wigner's Friend-thought experiment: when $\mathcal{W}$ performs a measurement in the $\brac{x}$-basis, after $\mathcal{F}$ has carried out a measurement in the $z$-basis, the state is found to be
\begin{equation*}
    \ket{\brac{x}_\mathcal{W}^+} = \frac{1}{\sqrt{2}}\left(\ket{f^+z_\mathcal{F}^+} + \ket{f^-z_\mathcal{F}^-}\right).
\end{equation*}
Thus, upon a subsequent measurement of $\brac{Z}_\mathcal{W}$, $\mathcal{W}$ should find either possible outcome with probability $\frac{1}{2}$. However, by \textbf{COO}, we surmise that whatever outcome $\mathcal{W}$ obtains will faithfully reveal $\mathcal{F}$'s actually observed outcome---thus, if $\mathcal{F}$ has in fact observed $v(Z_\mathcal{F}) = +1$, then $\mathcal{W}$ will obtain the outcome $\brac{z}_\mathcal{W}^+$ with certainty. 

The same conclusion follows for $\mathcal{B}$ directly from the definition of the eigenstates for $\brac{\brac{X}}_1$ and $\brac{\brac{Z}}_2$. In particular, in terms of $\mathcal{B}$'s observable $\mathcal{B}_1$, the state in Eq.~\eqref{BAS} is
\begin{equation}
    \ket{\Psi_\mathcal{BAS}} = \ket{\brac{\brac{z}}_1^+}\otimes\frac{1}{\sqrt{2}}\left(\ket{b_2^+\brac{z}_2^+} - \ket{b_2^-\brac{z}_2^-}\right).
\end{equation}
Now, presuming that $\mathcal{B}$ has obtained a definite outcome for their measurement of $\mathcal{B}_1$, we can appeal to \textbf{COO} and conclude that $\mathcal{C}$'s measurement must recapitulate this outcome; consequently, as was the case with $\mathcal{A}$, $\mathcal{C}$ could recover $\mathcal{B}$'s outcomes by measuring Friendified versions of $\mathcal{B}$'s observables. This again licenses $\mathcal{B}$ to conclude that there exist elements of reality associated with the $\mathcal{\brac{B}}_i$ with values $v\left(\mathcal{\brac{B}}_i\right)=v(\mathcal{B}_i)$.

Additionally, by \textbf{AR}, since $\mathcal{B}$ knows that $\mathcal{A}$ has carried out their measurements, and has concluded the existence of elements of reality for the Friendified versions of their observables, $\mathcal{B}$, too, is licensed to conclude that there are elements of reality corresponding to the $\mathcal{\brac{\brac{A}}}_i$ (although of course they cannot determine their values). Likewise, $\mathcal{A}$ knows that there must be elements of reality associated to the $\mathcal{\brac{B}}_i$ after $\mathcal{B}$ has carried out their measurements (and thus, before $\mathcal{C}$ performs theirs). In particular, this entails the existence of an element of reality for the product $\mathcal{\brac{\brac{A}}}_i\cdot\mathcal{\brac{B}}_i$, with
\begin{equation}
    v\left(\mathcal{\brac{\brac{A}}}_i\cdot\mathcal{\brac{B}}_i\right) = v(\mathcal{A}_i)\cdot v(\mathcal{B}_i).
\end{equation}

Consequently, $\mathcal{C}$ could simultaneously measure compatible, Friendified versions of $\mathcal{A}$'s and $\mathcal{B}$'s observables---i.e. $\brac{\brac{\mathcal{A}}}_i$ and $\brac{\mathcal{B}}_i$ for each row $i$. As this would recover the values observed by $\mathcal{A}$ and $\mathcal{B}$, and $\mathcal{C}_i = \brac{\brac{\mathcal{A}}}_i \cdot \brac{\mathcal{B}}_i$ (note that $\brac{\brac{X}}_2\cdot\brac{\brac{Z}}_2 = i\brac{\brac{Y}}_2$), we thus surmise that $v(\mathcal{C}_i) = v(\mathcal{A}_i)\cdot v(\mathcal{B}_i)$ for every row $i$.

Thus, \emph{before} any measurement is carried out by $\mathcal{C}$, but \emph{after} $\mathcal{A}$ and $\mathcal{B}$ have carried out their measurements, the following three propositions hold:
\begin{enumerate}
    \item If $\mathcal{C}$ measures $\mathcal{C}_1$, $v(\mathcal{C}_1)=v(\mathcal{A}_1)\cdot v(\mathcal{B}_1)$\label{C1}
    \item If $\mathcal{C}$ measures $\mathcal{C}_2$, $v(\mathcal{C}_2)=v(\mathcal{A}_2)\cdot v(\mathcal{B}_2)$\label{C2}
    \item If $\mathcal{C}$ measures $\mathcal{C}_3$, $v(\mathcal{C}_3)=v(\mathcal{A}_3)\cdot v(\mathcal{B}_3)$\label{C3}
\end{enumerate}

Consequently, after $\mathcal{A}$ and $\mathcal{B}$ have carried our their measurements, for any $i$, we could ask $\mathcal{C}$: ``What is the value of $v(\mathcal{A}_i)\cdot v(\mathcal{B}_i)$?", and, barring measurement errors, receive the correct answer with certainty. 

This situation can be compared to Schr\"odinger's example of the tired scholar \cite{schrodinger1935discussion}, who, faced with different questions in an examination, always gets at least the first one right---thus implying that they must know the answer to \emph{each} question beforehand, lacking any foreknowledge regarding which question would be asked (even if, upon being asked further questions, due to exhaustion, all following answers are essentially random).

Schr\"odinger's analogy was meant to illustrate the EPR argument \cite{einstein1935can}; there, of course, the `questions' being asked of the system are given by incompatible observables. Here, however, the observables $\mathcal{C}_i$ are perfectly compatible, and hence, simultaneously measurable. But there is a subtlety here that needs to be discussed.

Measuring, e.g., $\brac{\brac{\mathcal{A}}}_i$ and by extension $\mathcal{C}_i$ precludes simultaneously measuring $\brac{\mathcal{B}}_j$ for $j\neq i$. One might thus worry that, after measuring $\mathcal{C}_i$, assuming that $v(\mathcal{C}_j) = v(\mathcal{A}_j)\cdot v(\mathcal{B}_j)$ is no longer justified for $i\neq j$.

Suppose thus that $\mathcal{C}$ in fact measures $\mathcal{C}_1$. As $\mathcal{C}_1$ fails to commute, e.g., with $\brac{\brac{\mathcal{A}_2}}$, after this measurement, we can no longer argue that measuring $\brac{\brac{\mathcal{A}_2}}$ would recapitulate $\mathcal{A}$'s result of measuring $\mathcal{A}_2$. Hence, we can no longer appeal to the prior argumentation and conclude that since there must be elements of reality associated to $\brac{\brac{\mathcal{A}_2}}$ and $\brac{\mathcal{B}_2}$, it must be the case that $v(\mathcal{C}_2)=v(\mathcal{A}_2)\cdot v(\mathcal{B}_2)$.

However, we are of course not interested in the values of $\brac{\brac{\mathcal{A}_2}}$ and $\brac{\mathcal{B}_2}$, but in the value attached to their product; and there, it is not the case that measuring $\mathcal{C}_1$ leads to any incompatibility. Knowing the value of the product of two observables in incompatible contexts does not entail any knowledge of the value of either one. (See also the discussion of properties of composite systems in App.~\ref{subsec:comp}.) Hence, knowledge of $v(\mathcal{C}_i)$ does not in itself preclude knowledge of $v(\mathcal{C}_j)$, even if it precludes knowledge of $v(\mathcal{B}_j)$ or $v(\mathcal{A}_j)$. 

This can be made more clear by an explicit example. Suppose that initially, the system $\mathcal{S}$ is in the state
\begin{equation}
    \ket{\Psi_\mathcal{S}} = \ket{\Phi^+} = \frac{1}{\sqrt{2}}\left(\ket{z_1^+z_2^+} + \ket{z_1^-z_2^-}\right).
\end{equation}
Now, from Eq.~\eqref{BAS} and the surrounding discussion, we know that, after $\mathcal{A}$'s and $\mathcal{B}$'s measurements, in terms of $\mathcal{C}$'s observables, the system will be in the state
\begin{equation}\label{eq:doublebell}
    \ket{\Psi_\mathcal{BAS}} = \ket{\brac{\brac{\Phi^+}}} = \frac{1}{\sqrt{2}}\left(\ket{\brac{\brac{z}}_1^+\brac{\brac{z}}_2^+} + \ket{\brac{\brac{z}}_1^-\brac{\brac{z}}_2^-}\right),
\end{equation}
which is a simultaneous eigenstate of $\mathcal{C}$'s observables. In this state, by the above argumentation, we are justified, again, to conclude that for any $\mathcal{C}_i$, if $\mathcal{C}$ performs the corresponding measurement, then $v(\mathcal{C}_i) = v(\mathcal{A}_i)\cdot v(\mathcal{B}_i)$. But furthermore, all of $\mathcal{C}$'s observables have perfectly well-determined values prior to measurement---there are elements of reality associated with each $\mathcal{C}_i$. If a system is in an eigenstate of the measurements performed, then it is usually taken to be permissible, by the eigenstate-eigenvalue link, to talk of the values of the measured observables independently of their actual measurement---and it is \emph{of those preexisting values} that~\ref{C1}-\ref{C3} hold true.

The crucial point is thus that for each of those elements of reality, it must, before any measurement is made, be the case that $v(\mathcal{C}_i) = v(\mathcal{A}_i)\cdot v(\mathcal{B}_i)$; and furthermore, since $\mathcal{C}$'s measurements do not influence these elements of reality, even after, say, $\mathcal{C}_1$ has been measured, we are licensed to believe that, e.g., $v(\mathcal{C}_2) = v(\mathcal{A}_2)\cdot v(\mathcal{B}_2)$---even if we no longer have grounds to believe that $v\left(\brac{\brac{\mathcal{A}_2}}\right)=v(\mathcal{A}_2)$.

Note, again, that it is the (presumed) objectivity of $\mathcal{A}$'s (and $\mathcal{B}$'s) measurement result that warrants this inference. Given this objectivity, $\mathcal{C}$'s measurement will recapitulate the lower rung measurement results.

It is therefore no hindrance if the Friendified observables $\mathcal{\brac{\brac{A}}}_i$ and $\mathcal{\brac{B}}_i$ are never measured. In fact, this parallels an important part of the reasoning in the original Frauchiger-Renner experiment. In step~\ref{1}, $\mathcal{A}$ uses their own measurement result $\brac{x}_\mathcal{A}^+$ to conclude that the result of $\mathcal{F_B}$ must be equal to $z_\mathcal{B}^-$. But this is only true if the two-qubit system shared by $\mathcal{F_A}$ and $\mathcal{F_B}$ is in the state $\ket{x_\mathcal{A}^-z_\mathcal{B}^-}$; that is, $\mathcal{A}$ must infer the value of $X_\mathcal{A}$ from their observation of $\brac{X}_\mathcal{A}$, even though no measurement of $X_\mathcal{A}$ is ever performed. $\mathcal{A}$ must be able to, from their measurement, infer which component of the original Hardy-state obtains and thus, dictates the measurement outcome of $\mathcal{F_B}$; but this is equivalent to inferring that $v(X_A) = v(\brac{X}_A) = -1$, even though $X_A$ is never measured.

Assume now that the state of the system is given by the doubly-boxed Bell state $\ket{\brac{\brac{\Phi^+}}}$ as in Eq.~\eqref{eq:doublebell}. As we had surmised, \emph{in that state}, we have reason to believe that $v(\mathcal{C}_i) = v(\mathcal{A}_i)\cdot v(\mathcal{B}_i)$ for every row $i$. Thus, we can now run the argument in Sec.~\ref{subseq:PMW} with this starting state, avoiding the worry about measurement incompatibilities.

Suppose now that $\mathcal{C}$ records their having obtained a contradiction in a quantum register (which we may take to be a single qubit) by putting it in the state $\ket{\lightning_\mathcal{C}}$. So the evolution of that particular degree of freedom in the above scenario will be
\begin{equation}
    \ket{?_\mathcal{C}}\otimes\ket{\brac{\brac{\Phi^+}}}\leadsto \ket{\lightning_\mathcal{C}}\otimes\ket{\brac{\brac{\Phi^+}}}.
\end{equation}

But the same argumentation as above is possible for the other doubly-Friendified Bell states, $\ket{\brac{\brac{\Phi^-}}}$, $\ket{\brac{\brac{\Psi^+}}}$ and $\ket{\brac{\brac{\Psi^-}}}$. But since these form a basis for $\mathcal{C}$'s system, by linearity, for an arbitrary state $\ket{\Psi_\mathcal{BAS}}$ we have (where $\alpha$, $\beta$, $\gamma$ and $\delta$ are normalization coefficients):
\begin{align}
    \ket{?_\mathcal{C}}&\otimes\ket{\Psi_\mathcal{BAS}} = \ket{?_\mathcal{C}}\otimes\left(\alpha\ket{\brac{\brac{\Phi^+}}} + \beta\ket{\brac{\brac{\Phi^-}}} + \gamma\ket{\brac{\brac{\Psi^+}}} + \delta\ket{\brac{\brac{\Psi^-}}}\right) \\
    & \nonumber= \alpha\ket{?_\mathcal{C}}\otimes\ket{\brac{\brac{\Phi^+}}} + \beta\ket{?_\mathcal{C}}\otimes\ket{\brac{\brac{\Phi^-}}} + \gamma\ket{?_\mathcal{C}}\otimes\ket{\brac{\brac{\Psi^+}}} + \delta\ket{?_\mathcal{C}}\otimes\ket{\brac{\brac{\Psi^-}}} \\
    & \nonumber\leadsto \alpha\ket{\lightning_\mathcal{C}}\otimes\ket{\brac{\brac{\Phi^+}}} + \beta\ket{\lightning_\mathcal{C}}\otimes\ket{\brac{\brac{\Phi^-}}} + \gamma\ket{\lightning_\mathcal{C}}\otimes\ket{\brac{\brac{\Psi^+}}} + \delta\ket{\lightning_\mathcal{C}}\otimes\ket{\brac{\brac{\Psi^-}}} \\
    & \nonumber = \ket{\lightning_\mathcal{C}}\otimes \ket{\Psi_\mathcal{BAS}}
\end{align}

Thus, $\mathcal{C}$ will encounter a contradiction for every possible starting state. Consequently, no matter what state the system in $\mathcal{A}$'s lab is in, in the given scenario, it is not possible for $\mathcal{A}$, $\mathcal{B}$, and $\mathcal{C}$ to have a mutually consistent experience: the Rashomon effect is generic in the quantum world.

\begin{figure}[h] 
 \centering
 \begin{overpic}[width=\textwidth]{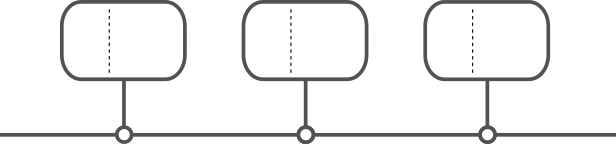}
  \put(13,20){$\mathcal{A}_1$}
  \put(19,20){$\mathcal{B}_1=\mathcal{C}_1$?}
  \put(13,16){$\mathcal{A}_2$}
  \put(19,16){$\mathcal{B}_2=\mathcal{C}_2$?}
  \put(13,12){$\mathcal{A}_3$}
  \put(19,12){$\mathcal{B}_3=\mathcal{C}_3$?}
  \put(42,20){$\mathcal{B}_1$}
  \put(48,20){$\mathcal{A}_1=\mathcal{C}_1$?}
  \put(42,16){$\mathcal{B}_2$}
  \put(48,16){$\mathcal{A}_2=\mathcal{C}_2$?}
  \put(42,12){$\mathcal{B}_3$}
  \put(48,12){$\mathcal{A}_3=\mathcal{C}_3$?}
  \put(72,20){$\mathcal{C}_1$}
  \put(77.5,20){$\mathcal{A}_1=\mathcal{B}_1$?}
  \put(72,16){$\mathcal{C}_2$}
  \put(77.5,16){$\mathcal{A}_2=\mathcal{B}_2$?}
  \put(72,12){$\mathcal{C}_3$}
  \put(77.5,12){$\mathcal{A}_3=\mathcal{B}_3$?}
  \put(19,-2){$t_1$}
  \put(48.5,-2){$t_2$}
  \put(78,-2){$t_3$}
\end{overpic}
\caption{A timeline of the proposed experiment, indicating the obtained facts for each observer at each point in time.}
\label{pic:time}
\end{figure}

\begin{objection}[Influence of Higher-Level Measurements]
    In general, $\mathcal{B}$'s measurements will not leave $\mathcal{A}$'s state invariant. Is it then still permissible to talk of $\mathcal{A}$'s, $\mathcal{B}$'s, and $\mathcal{C}$'s values as simultaneous elements of reality?
\end{objection}

Quantum measurement generally entails a change of state of the measured system. Consequently, we must suppose that, in general, after a higher-rung measurement, information at the lower rung is not preserved. And indeed, a state like $\ket{\brac{\brac{z}}_1^+\brac{\brac{x}}_2^-}$ is not a state in which $\mathcal{A}$ has observed the outcomes $\mathcal{A}_1 = +1$, $\mathcal{A}_2 = -1$, and $\mathcal{A}_3 = -1$---or, in fact, is in a state of having observed any definite outcome at all. But this still does not allow us to tell a consistent story encompassing $\mathcal{A}$'s, $\mathcal{B}$'s, and $\mathcal{C}$'s observations. To see this, consider Fig.~\ref{pic:time}. There, the temporal sequence of the experiment is made explicit, together with the knowledge of each observer at each point in time. Thus, $\mathcal{A}$ knows the value of their observables after having measured them---however, crucially, this lets them also make predictions of the observations made by $\mathcal{B}$ and $\mathcal{C}$, \emph{at the time they make them}. This is due to the fact that $\mathcal{A}$, at $t_1$, making the $Z_1$-measurement knows that, due to \textbf{COO}, $\mathcal{C}$, at $t_3$, will obtain the same value for $\brac{\brac{Z}}_1$.

$\mathcal{A}$, at $t_1$ due to their observations, may thus predict whether observed values of $\mathcal{B}$ and $\mathcal{C}$ agree or disagree. Likewise, at $t_2$ and $t_3$, each of $\mathcal{B}$ and $\mathcal{C}$ may analogously judge whether the observables of the other two observers agree in the corresponding rows. However, as the above discussion implies, there is no sequence such that these judgements are mutually consistent. This does not depend on whether each observer's experience is undone by a higher-rung measurement---whether, in this sense, they forget their predictions after making them.

$\mathcal{A}$ is thus in no worse shape than $\mathcal{F_B}$ in the original Frauchiger-Renner argument: $\mathcal{B}$'s measurement will undo their observations---after final projection on the $\ket{\brac{x}_\mathcal{A}^-\brac{x}_\mathcal{B}^-}$-component, $\mathcal{F_B}$ will not be in a state where they have made a definite observation of $v(Z_\mathcal{A})$---yet their predictions still hold fast. $\mathcal{A}$ knows, after their measurement, the state to be (say) $\ket{z_1^+x_1^-}$; therefore, they can conclude that in the $\brac{\brac{z}}_1\brac{\brac{x}}_2$-basis, the state will be $\ket{\brac{\brac{z}}_1^+\brac{\brac{x}}_1^-}$, and consequently, a measurement of $\brac{\brac{Z}}_1$ and $\brac{\brac{X}}_2$ will replicate their results. Likewise, $\mathcal{F_B}$ knows, after their measurement, the state to be $\ket{z_\mathcal{A}^-z_\mathcal{B}^-}$, and thus, that $\mathcal{F_A}$ must necessarily obtain the $-1$-outcome for their measurement.

Hence, we cannot associate a consistent story covering all observer's experiences simultaneously. Consequently, the argument's conclusion cannot be alleviated by appealing to the operations performed by higher-rung observers on lower-rung observers: to the extent that it is reasonable to speak of each as having made definite observations, these have independent validity, yet are mutually inconsistent.

\begin{objection}[Existence of Hidden Variable-Accounts]
    The argument purports to show that no single reality---that is, no consistent simultaneous assignment of values to all observables---is possible. But in hidden variable-interpretations of quantum mechanics, e.g. the de Broglie-Bohm formulation \cite{deB1927, bohm1952}, there always exists a single assignment of values. Is there a `Rashomon-effect' in this case?
\end{objection}

The aim of the original Peres-Mermin argument is to demonstrate the Kochen-Specker theorem \cite{KS1967}. Thus, if there are values associated to all observables in an experiment over and above those predicted by the quantum formalism (hidden variables), in general, this assignment must be \textit{contextual}: that is, the value of, say, $\mathcal{A}_1$ measured together with $\mathcal{A}_2$ and $\mathcal{A}_3$ cannot be assumed to be the same as if it is measured together with $\mathcal{B}_1$ and $\mathcal{C}_1$. 

Translated to the present situation, this means that $\mathcal{A}$, having measured all of their observables, cannot appeal to \textbf{COO} to obtain information about $v(\mathcal{B}_i)$ and $v(\mathcal{C}_i)$: the values they observed cannot be the same as those considered in the context of the measurements implemented by $\mathcal{B}$ and $\mathcal{C}$. The contradiction derived above entails that one of the assumptions in its derivation must be false. The existence of a single assignment of values to all observables in the case of a hidden variable-interpretation then comes at the expense of violating \textbf{COO}: the assignment is determined by the context within which it is considered. This is a `single reality' in the sense of there being such an assignment at all times; however, it cannot be considered a unified reality for all of $\mathcal{A}$, $\mathcal{B}$ and $\mathcal{C}$. In this sense, it seems apt to talk of a Rashomon effect in this case, as well.

\section{Epistemic Horizons and the Rashomon Effect} \label{epihor}

The above seems to show that, relying on the predictions of quantum theory, we may arrive at contradictory conclusions regarding the outcome of certain experiments. However, there is room for a different view, which makes use of the notion of \emph{epistemic horizons}, introduced in \cite{szangolies2018epistemic} (see also \cite{szangolies2020epi}, \cite{poletti2025observable} and \cite{fankhauser2025epistemic} for further and related developments).

\subsection{Epistemic Horizons in the Frauchiger-Renner Argument}

Frauchiger and Renner intend their argument to show the inconsistency of the application of quantum theory to its own use; but this is not the only possible conclusion. Rather, we may surmise that it entails a quantum version of the Rashomon effect: the existence of narratives that cannot be unified into a single, all-encompassing, consistent meta-narrative. 

To see this, consider the information content in the state after $\mathcal{A}$'s measurement in step~\ref{1}. According to the concept of an epistemic horizon (see Appendix~\ref{App:EpiHor} for details), there exists a maximum amount of information available to any given system about another system---in the case of a two-qubit system, two bits of information are maximally available. Thus, $\mathcal{A}$ can at most answer two distinct elementary questions (having a yes/no answer) about the system. 

After measurement of $X_\mathcal{A}$, one bit of available information is given by the obtained result, which in accordance with the argument we will suppose to be $x_\mathcal{A}^+$. (We suppress the level of Friendification here, as we take it as given that, by \textbf{COO}, observations of Friendified observables imply the value of non-Friendified analogues.) But the state (which $\mathcal{A}$ knows) of the system is such that $\mathcal{A}$ may derive a further bit of knowledge: namely, the correlation between $X_\mathcal{A}$ and $Z_\mathcal{B}$ as given by the Hardy state in Eq.~\eqref{xz}. 

For such situations, the framework of epistemic horizons introduces the notion of a \emph{conditional event}: that is, a value of an observable that is only definite conditional to the value of another observable---as is the case in the Hardy state: only conditioned on the value of $X_\mathcal{A}$ being $x_\mathcal{A}^-$ does $Z_\mathcal{B}$ possess the value $z_\mathcal{B}^-$ (for more details, see Appendix~\ref{subsec:comp} and Ref.~\cite{szangolies2020epi}). We denote this as $z_\mathcal{B}^-|x_\mathcal{A}^-$ (read: `$z_\mathcal{B}=-1$ given that $x_\mathcal{A}=-1$'), and thus, can write the total information content in the state of the system as available to $\mathcal{A}$ as $(x_\mathcal{A}^-,z_\mathcal{B}^-|x_\mathcal{A}^-)$. 

The important realization here is that such a state does not allow us to inquire into what \emph{would} have been the case, \emph{had} $\mathcal{A}$ not made the observation of $x_\mathcal{A}^-$---because it is only conditional on this observation that we can speak of the value of $Z_\mathcal{B}$ as being $z_\mathcal{B}^-$. Thus, knowing $z_\mathcal{B}^-|x_\mathcal{A}^-$ is distinct from knowing $z_\mathcal{B}^-$ simpliciter: while it is equivalent as regards experimental predictions (in both cases, $\mathcal{A}$ may predict that a measurement of $Z_\mathcal{B}$ yields, in fact, $z_\mathcal{B}^-$), it differs regarding the counterfactuals that are supported by the available information.

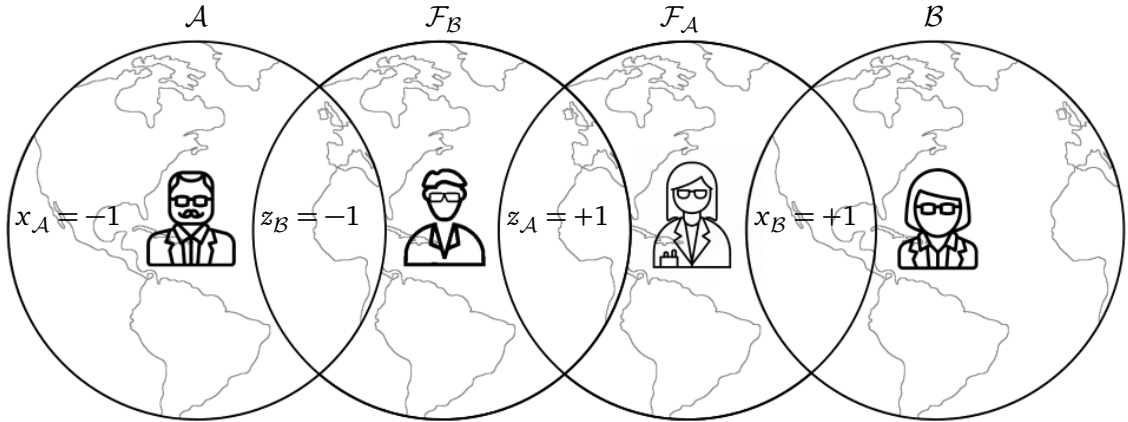
\begin{figure}[h] 
 \centering
 \begin{overpic}[width=\textwidth]{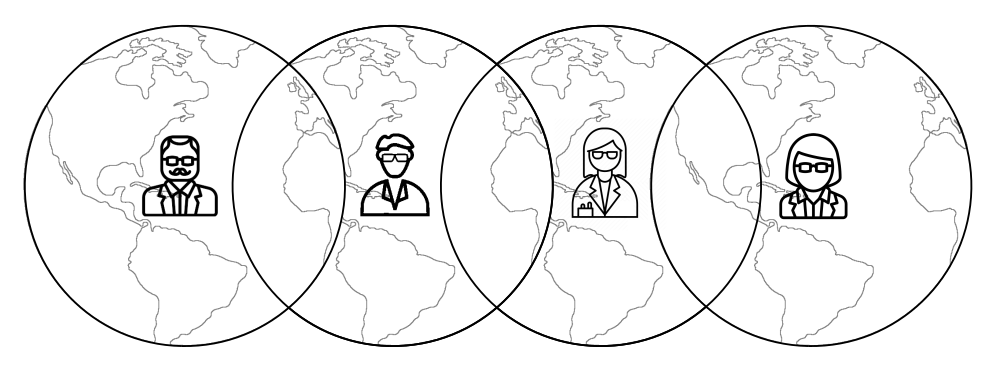}
  \put(17.5,36){$\mathcal{A}$}
  \put(38,36){$\mathcal{F_B}$}
  \put(58,36){$\mathcal{F_A}$}
  \put(80.5,36){$\mathcal{B}$}
  \put(3,19){$x_\mathcal{A}=-1$}
  \put(24,19){$z_\mathcal{B}=-1$}
  \put(45,19){$z_\mathcal{A}=+1$}
  \put(66,19){$x_\mathcal{B}=+1$}
\end{overpic}
\caption{Epistemic horizons of the observers in the Frauchiger-Renner argument: while horizons may overlap, they cannot be integrated into an overarching `meta-horizon'.}
\label{pic:hori}
\end{figure}

The argument then proceeds by using $z_\mathcal{B}^-$ to conclude that $\mathcal{F_A}$ must obtain the $+1$-outcome; but this is not a conclusion $\mathcal{A}$ can draw: in doing so, they would exceed the maximum information available about a system. Knowledge that $z_\mathcal{B}^-|x_\mathcal{A}^-$ cannot stand in for knowledge that $z_\mathcal{B}^-$. Each observer must, in drawing inferences about other observations, remain mindful of the limit of their epistemic horizon (see Fig.~\ref{pic:hori}).

Information from within different epistemic horizons cannot be integrated into a single, coherent narrative: doing so would entail exceeding the information limit. Hence, we should not be surprised that the attempt to do so invites inconsistency. 

We can compare this to the example of a one-time pad. There, any bit of the key is added to a bit of the cipher to obtain the clear text; consequently, the value of one bit of the cipher, and one bit of the key, entails the corresponding bit of the message. Knowledge of $z_\mathcal{B}^-|x_\mathcal{A}^-$ is akin to knowing one bit of the key: only in conjunction with knowledge of $x_\mathcal{A}^-$---the cipher text---does it license any conclusions about $Z_\mathcal{B}$. 

Think of $\mathcal{A}$ as equipped with a two-bit memory register to store knowledge about the system: we start out with it containing $x_\mathcal{A}^-$ and $z_\mathcal{B}^-|x_\mathcal{A}^-$---which jointly entail $z_\mathcal{B}^-$. But in order to now use $z_\mathcal{A}^+|z_\mathcal{B}^-$ to derive any further conclusions (step~\ref{2} of the argument), $\mathcal{A}$ would have to purge one bit of information from their register. If $\mathcal{A}$ purges $z_\mathcal{B}^-|x_\mathcal{A}^-$, nothing follows about $Z_\mathcal{A}$ from $x_\mathcal{A}^-$ and $z_\mathcal{A}^+|z_\mathcal{B}^-$; but the same is true if $\mathcal{A}$ keeps it, purging $x_\mathcal{A}^-$ instead: $z_\mathcal{B}^-|x_\mathcal{A}^-$ and $z_\mathcal{A}^+|z_\mathcal{B}^-$---keys to two different ciphers---do not jointly carry any information about $Z_\mathcal{A}$.

\subsection{Epistemic Horizons in the Peres-Mermin-Wigner Argument}

The same general conclusion then holds true for the three-agent argument presented here. $\mathcal{C}$, after their measurement, knows the values of the propositions $\mathcal{C}_1 \equiv \mathcal{A}_1|\mathcal{B}_1$ and $\mathcal{C}_2 \equiv \mathcal{A}_2|\mathcal{B}_2$ (and by extension, $\mathcal{C}_3 \equiv \mathcal{A}_3|\mathcal{B}_3$)---that is, whether $v(\mathcal{A}_i) = v(\mathcal{B}_i)$ for every row $i$. But these items of knowledge, under the epistemic horizon-framework, do not support counterfactuals regarding $\mathcal{A}$'s and $\mathcal{B}$'s values, as to engage in such counterfactual reasoning would exceed the boundaries on maximally accessible information. 

\begin{figure}[h] 
 \centering
 \begin{overpic}[width=.7\textwidth]{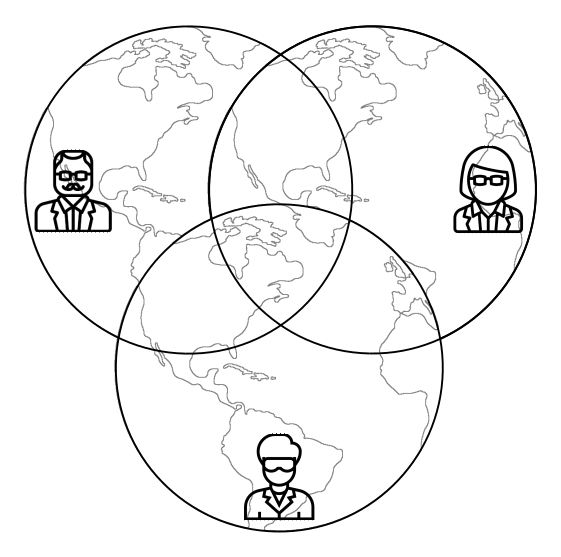}
    \put(7,90){$\mathcal{A}$}
    \put(88,90){$\mathcal{B}$}
    \put(48,0){$\mathcal{C}$}
    \put(22,68){$\mathcal{B}_i = \mathcal{C}_i$}
    \put(66,68){$\mathcal{A}_i = \mathcal{C}_i$}
    \put(43,25){$\mathcal{A}_i = \mathcal{B}_i$}
    \put(42,71){$(\mathcal{A}_i\cdot \mathcal{B}_i)|\mathcal{C}_i$}
    \put(27,44){$(\mathcal{A}_i\cdot \mathcal{C}_i)|\mathcal{B}_i$}
    \put(59,44){$(\mathcal{B}_i\cdot \mathcal{C}_i)|\mathcal{A}_i$}
    \put(43,53){$\mathcal{A}_i\cdot \mathcal{B}_i\cdot \mathcal{C}_i$}
\end{overpic}
\caption{Epistemic horizons of the observers in the Peres-Mermin-Wigner argument. Note that here, we are referring to the truth values of propositions: that is, $\mathcal{A}$ knows, e.g., whether $\mathcal{B}_1 = \mathcal{C}_1$, while both $\mathcal{A}$ and $\mathcal{B}$ know the value of $\mathcal{A}_1 \cdot \mathcal{B}_1$, conditional on knowing $\mathcal{C}_1$.}
\label{pic:pmwhor}
\end{figure}

$\mathcal{C}$ can only attach values to $\mathcal{A}$'s observables upon stipulating values obtained by $\mathcal{B}$; but this requires exactly the sort of counterfactual reasoning not supported by the conditional items of knowledge they have access to. While thus $\mathcal{C}$ may stipulate that the values obtained by $\mathcal{A}$ and $\mathcal{B}$ in the first row must agree, any further reasoning of the form `if $\mathcal{B}$ has obtained the value $+1$ for $\mathcal{B}_1$, then...' is precluded by the epistemic horizon.

This is illustrated in Fig.~\ref{pic:pmwhor}. Again, $\mathcal{C}$ can only make determinations on the values of $\mathcal{A}$ conditional on those of $\mathcal{B}$. But stipulating this information entails `overstepping' the boundaries of $\mathcal{C}$'s epistemic horizon: $\mathcal{C}$'s two bits of knowledge are exhausted by their observed outcomes, thus, appealing to additional information about the system goes beyond this maximum. 

In the analogy from the previous section, $\mathcal{C}$'s two-bit register, after their measurement, contains information on the correlations between $\mathcal{A}$'s and $\mathcal{B}$'s outcomes, and only those correlations. Thus, appealing to any given actual value of either would entail erasing one such bit. Consequently, if $\mathcal{C}$ were to appeal to, say, $\mathcal{B}_1 = 1$, and use their result of $\mathcal{C}_1 = -1$ to conclude that $\mathcal{A}_1 = -1$, to `store' this information, that about the correlation between $\mathcal{A}_2$ and $\mathcal{B}_2$ would have to be erased---and consequently, the argument could not proceed.

This view thus restricts the applicability of \textbf{COO} to consistency relations that can be checked within a given horizon, as determined by the maximum information available to an observer. In this sense, quantum mechanics does not fail to consistently describe the use of itself; rather, in a quantum world, not all narratives can be integrated into a single, overarching meta-narrative. In contrast to the classical world, distinct observers may thus have experiences that cannot be brought into a mutually consistent framework, without either of them being wrong: in the quantum world, the Rashomon effect is not just a mere artifact of unreliable narration.

A related view, championed e.g. in \cite{brukner2018no,kochen2015reconstruction,bub2020understanding}, is that there is no single Boolean algebra encompassing all observations. In this case, the problematic inferences are blocked by the nonexistence of a single (Boolean) lattice encompassing all possible propositions relating to measurement outcomes. In contrast, the epistemic horizon takes an informational approach, putting a bound on the maximum information available about a system to any given observer, itself derived by means of an independent argument in Ref.~\cite{szangolies2018epistemic}. 

It would be interesting, however, to investigate the connection between the two approaches by means of e.g. the notion of \textit{Boolean frames} as introduced in Ref.~\cite{cuffaro2023measurement}. There, a Boolean frame refers to a maximal Boolean subalgebra of the lattice of projections in Hilbert space representing the simultaneously answerable yes-or-no questions associated to a given measurement context. In this sense, Boolean frames may serve as a formalization of epistemic horizons within the observable algebra of quantum mechanics.

Another similar approach, originating within the `consistent histories'-approach to quantum mechanics, is offered in Ref.~\cite{losada2019frauchiger}. There, it is shown that the Frauchiger-Renner argument appeals to a family of histories that fails to obey the consistency condition to achieve its contradiction; however, the relevance of this observation to the argument has been questioned \cite{bub2020understanding}.

Relatedly the notion of a `Wigner bubble' was proposed in Ref.~\cite{cavalcanti2021view} within a QBist perspective. From the point of view of the pragmatist approach characteristic of QBism, events themselves are relativized to a given observational context, thus failing to be absolute. In contrast, while both the Wigner bubble and the epistemic horizon each center their narratives on the perspective of a given observer, the latter merely prohibits certain \textit{conclusions about} events beyond the horizon.

A common thread among these approaches is to limit the set of simultaneously valid inferences to some contradiction-free subset. However, most of these take quantum mechanics as a starting point, \textit{deriving} the limitation on simultaneous information only as a secondary matter. In contrast, epistemic horizons form a \textit{prior} restriction on the accessible information for any kind of observer, with characteristically quantum features appearing as a consequence~\cite{szangolies2018epistemic}. 

\section{Conclusion} \label{conc}

The aim of the preceding has been two-fold: to propose a novel Wigner's Friend-type setting substantially strengthening the conclusions of the Frauchiger-Renner argument, and to interpret the resulting picture in terms of a Rashomon effect---the impossibility of integrating multiple perspectives into a coherent whole---brought about by the limitation in information available about a system due to the existence of epistemic horizons.

To this effect, it has been proposed that, rather than a failure of quantum mechanics to `consistently describe the use of itself', the apparent contradiction in Frauchiger-Renner-like arguments obtains only if we attempt to integrate these distinct narratives or perspectives into a single, overarching whole---and thus, highlights nothing but the impossibility of doing so. In a classical world, such a circumstance is psychological, owing to the unreliable nature of memory, or lacking factual accuracy of the narrative---there is always one true story, which inconsistent testimony merely recounts imperfectly. Frauchiger and Renner cast this in terms of their Assumption \textbf{C}, requiring that `a theory T [...] must allow any agent $\mathcal{A}$ to promote the conclusions drawn by another agent $\mathcal{A}^\prime$ to his own conclusions' \cite{frauchiger2018quantum}.

But in the quantum world, things may not be so simple. If, as previously proposed \cite{szangolies2018epistemic,szangolies2020epi}, there exists a fundamental limit to the information available about any given system, then the sort of reasoning necessary to carry out the inferences involved in the Frauchiger-Renner argument by the involved parties runs afoul of this boundary: we cannot peek behind the epistemic horizon.

This sort of `quantum perspectivism' is not entirely new. In the consistent histories framework, it is enshrined in the `single framework'-rule \cite{griffiths2014new}; Bub has emphasized that there is no `view from nowhere' in quantum mechanics \cite{bub2020understanding}; Kochen has provided a reconstruction of quantum mechanics in terms of `$\sigma$-complexes' comprising the union of the $\sigma$-algebras associated to different measurement contexts \cite{kochen2015reconstruction}; the relational interpretation of Rovelli et al. relativizes the concept of `fact' to a particular observer \cite{rovelli1996relational}; Cavalcanti's `Wigner bubbles' does similar with the notion of an event \cite{cavalcanti2021view}; and so on. What is new in the framework of epistemic horizons is, rather, the attempt to localize the origin of the fractionalization of the description of the world into partially disjoint perspectives in intrinsic limitations on the availability of information about a system.

For a two-qubit system, then, any given observer may obtain at most two bits of information about its state. In a composite system in which that information is solely taken up by correlations, as opposed to the values of any of its constituent part's properties---that is, an entangled system---measurements on one of its subsystems yield conditional knowledge about the other. Thus, if $\mathcal{A}$ measures $x_\mathcal{A}$ in such a system, their knowledge of the distant part's $x$-property will be given by what we have called a \emph{conditional event}: obtaining, say, the $-1$-value, the resulting state can be written as $(x_\mathcal{A}^-,x_\mathcal{B}^+|x_\mathcal{A}^-)$. This exhausts the two bits of information available to $\mathcal{A}$.

The important thing to realize is now that $\mathcal{A}$'s knowledge in this case differs from that in a state such as $(x_\mathcal{A}^-,x_\mathcal{B}^+)$: in both cases, $\mathcal{A}$ may confidently predict that $\mathcal{B}$, upon carrying out an $x$-measurement, will observe the $-1$-outcome; but in the former case, they can do so only \emph{conditionally on the fact that they have, in fact, obtained the $-1$ outcome for their $x$-measurement}. In the latter case, $\mathcal{A}$ may reason that even if they had obtained the $+1$-value, or made a different measurement altogether, $\mathcal{B}$ would have obtained the same result. 

Consequently, knowledge of $x_\mathcal{B}^+|x_\mathcal{A}^-$ does not license the same inferences as knowledge of $x_\mathcal{B}^+$ simpliciter does. In particular, $x_\mathcal{B}^+|x_\mathcal{A}^-$ cannot stand for $x_\mathcal{B}^+$ in further inferences: thus, if we had, for instance, another conditional event of the form $z_\mathcal{C}^+|x_\mathcal{B}^+$, $\mathcal{A}$ cannot derive $x_\mathcal{B}^+$ from $x_\mathcal{B}^+|x_\mathcal{A}^-$ and $x_\mathcal{A}^-$, and then $z_\mathcal{C}^+$ from $z_\mathcal{C}^+|x_\mathcal{B}^+$ and $x_\mathcal{B}^+$. Doing so, they would have access to \emph{three} bits of information about the state of the system---but only two bits are consistently attainable. All further information must remain behind their epistemic horizon.

Thus, the picture in the Frauchiger-Renner scenario is that of four observers $\mathcal{A}$, $\mathcal{B}$, $\mathcal{F_A}$ and $\mathcal{F_B}$, each of which may have access to at most two bits of information about their shared quantum state---but not necessarily \emph{the same} two bits. For this to be possible, each of these observers must, in a sense, inhabit their own `world'---or `reality', if we go by the definition of reality implicit in the term `local realism' as used in Bell's theorem. This, then, leads to the idea that in the quantum world, the Rashomon effect is not merely a psychological, but a fundamental notion: there really are mutually irreconcilable experiences of the same circumstance.

However, the original Frauchiger-Renner argument suffers from some potential shortcomings. First of all, it is both probabilistic, and depends on a special choice of initial state, necessitating entanglement between spatially separate systems---indeed, even between the laboratories of $\mathcal{F_A}$ and $\mathcal{F_B}$ as a whole. Furthermore, to obtain a definite contradiction, the measurement dynamics need to be invoked in the end, leading to the collapse of the state onto the $\ket{\brac{x}_\mathcal{A}^-\brac{x}_\mathcal{B}^-}$-component. This non-unitary change, one might hold, `undoes' the reality of what went on before, generally leading to a loss of information. On these grounds, one might seek to resist the idea of irreconcilable experiences.

But this view seems much harder to maintain with the proposed argument based on the Peres-Mermin square of observables. As the contradiction follows deterministically, it does not depend on any final application of the projection postulate. Furthermore, at any given point during or after the experiment, $\mathcal{C}$ might prompt each of $\mathcal{A}$ or $\mathcal{B}$ to indicate whether they have observed some definite outcome, and each will agree; consequently, at each point, $\mathcal{C}$ will be able to deduce that whatever values each has observed must be inconsistent with $\mathcal{C}$'s. 

Indeed, one may even propose that the system containing the three observers remains its own, closed-off universe: then, within this universe, there will be definite facts of the matter that each of the parties has made some definite observation, and that these experiences are mutually irreconcilable.

Thus, the proposed thought experiment---barring additional dynamics, such as a spontaneous collapse---seems to provide strong support for the notion that the Rashomon effect, the existence of incongruous narratives, is a fundamental phenomenon within the framework of quantum mechanics. Each of the observers in the scenario has their own, definite experience; but every other observer's experience remains forever closed off behind their respective epistemic horizons.

The original aim of the Frauchiger-Renner argument was to show that `single-world interpretations of quantum theory cannot be self-consistent' \cite{frauchiger2016single}. But what, ultimately, is a world? According to the famous dictum of Wittgenstein, `the world is the totality of facts, not of things' \cite[prop.~1.1]{wittgenstein1922tractatus}. The above then may be taken to constitute grounds for believing that no such totality exists: in that sense, one might be inclined to agree with the original sentiment---not necessarily because there must be a multiplicity of worlds, but because there cannot be even a single, universally shared one.

This failure can be taken in two distinct senses, with no firm commitment to either entailed by the idea of epistemic horizons itself. First, one can take this `fracturing' of reality in an ontic sense---there really are distinct, only partially overlapping realities out there, which are faithfully captured by the quantum formalism. This approach might be associated with the consistent histories framework \cite{griffiths2014new}, or the relational interpretation \cite{rovelli1996relational}. The second is, then, its epistemic counterpart---stipulating that while there might be a single reality out there, all we have access to is a collection of incomplete, partial, yet jointly necessary descriptions thereof. On this latter sort of view, perhaps most closely related to QBism \cite{fuchs2014introduction} or the informational interpretation of Bub and Pitowsky \cite{bub2010two,bub2020two}, the world outruns our ability to create descriptions of it---there is no all-encompassing model of the `whole world', in itself, no complete atlas of its geography, but merely many local maps, patched together, and overlapping only partially.

In the last consequence, this comes down to a difference in attitude towards those measurement results for which the epistemic horizon prohibits the association of a definite value---with the ontic view holding that therefore, these values do not exist, and the epistemic view asserting that they simply cannot be predicted. While no conclusive decision between these alternatives is possible on the basis of the preceding arguments, the more modest approach might be not to expect Mother Nature to be constrained by the limits of our perspective.

\section*{Acknowledgements}
I thank Michael Epping and two anonymous referees of an earlier version of this article for their insights and suggestions.

\section*{Declarations}

\textbf{Funding.} This research received no external funding.

\noindent\textbf{Clinical Trial Number.} Not applicable

\noindent\textbf{Consent to Publish.} Not applicable

\noindent\textbf{Ethics and Consent to Participate.} Not applicable

\section{Data Availability Statement}
No data were produced during the compilation of this study. In case of questions, please contact the author.

\begin{appendix}

\section{Fundamentals of Epistemic Horizons}\label{App:EpiHor}
The starting point for the notion of an epistemic horizon is the observation that several reconstructions of quantum mechanics starting from information-theoretical notions make use of two fundamental principles \cite{grinbaum2003elements}:

\begin{enumerate}[label=\Roman*]
 \item \label{fin}\textit{Finiteness}: There is a finite maximum of information that can be obtained about any given system. 
 \item \label{add}\textit{Extensibility}: It is always possible to acquire new information about any system.
\end{enumerate}

These directly entail that obtaining new information about a system, in order not to violate the bound proposed by principle~\ref{fin}, must invalidate prior information---that is, measurements generically lead to state-changes. Moreover, they lead to a maximum `resolution' of phase-space, that is, a smallest phase-space volume within which any given system can be localized---which introduces the constant $\hbar$ as a measure of this volume, and the uncertainty relation as an expression of this localizability (cf. \cite{rovelli1996relational}). For a more thorough discussion, and an explicit derivation of the finiteness of information obtainable about a system based on algorithmic incomputability see Ref.~\cite{szangolies2018epistemic}, which also includes illustrations of how familiar quantum phenomena emerge from the above two constraints.

The two principles~\ref{fin} and \ref{add} may be considered as having the same status as the principle of relativity and the finiteness of the speed of light in the derivation of special relativity, and simply be postulated in an axiomatic way. However, in \cite{szangolies2018epistemic}, it was argued that they can in fact be derived from a more fundamental argument. To fix some notation, let $\mathcal{S}$ be a system that may be in countably many different states $\{s_i\}_{i\in\mathbb{N}}$ (although countability is not necessary \cite{szangolies2018epistemic}, it simplifies the formulation). Furthermore, assume that there are dichotomic measurements $\{m_j\}_{j\in\mathbb{N}}$, such that $m_j(s_i)\in\{-1,1\}$.

Then, let us formulate the following \emph{principle of classicality}:

\begin{theorem}[Classicality]
\label{assump:class}
For every state $s_k$ and measurement $m_n$, there exists a function $f$ such that $f(n,k)=m_n(s_k)$.
\end{theorem}

This captures the intuition that, classically, one ought to be able to gather sufficient information about a system such that the result of every further measurement is deducible from this information (say, by means of computation).

It can then be shown that this assumption leads to a contradiction; and that moreover, only finitely many values of $f(n,k)$ can be determined for any given state $s_k$ (for details, see \cite{szangolies2018epistemic}; for a related argument, see \cite{landsman2020indeterminism}). This entails principle~\ref{fin}.

Every measurement can be mapped to the characteristic function for a region of phase space. Thus, by appealing to the algebra of subsets of phase space, we can build a logical calculus of measurements (or equivalently, properties the system possesses). From two measurements $m_1$ and $m_2$, we can then construct a third, $m_{12} = m_1 \oplus m_2$, where the operator `$\oplus$' denotes the binary \texttt{xor}: $m_{12} = 1$ if $m_1 = m_2$, and $m_{12} = -1$ if $m_1 \neq m_2$.

For every measurement thus constructed, an explicit measuring procedure can be given: simply measure position and momentum up to the required degree of precision to localize the system within the given subset. But then, measurement procedures can be given for measurements whose values are not given by $f(n,k)$; hence, information not already possessed about a system can be obtained. Thus follows principle~\ref{add} \footnote{One might suppose that certain measurements simply produce no outcome, in whatever way, to try and evade the dilemma. However, this does not produce a viable escape: we can simply formulate an analogous argument for the three outcomes $\{-1,1,\emptyset\}$, where `$\emptyset$' denotes `no outcome' or `ambiguous' \cite{szangolies2018epistemic}.}. 

\subsection{Elementary Systems}

The simplest conceivable system, in the above formalism, is a system $\mathcal{S}$ for which $f(n,k)$ produces just a single definite value. This then implies that there must be at least one indefinite property. A system with two dichotomic properties may be in one of four possible states (say $\{00, 01, 10, 11\}$), and knowledge of one of them---corresponding to having one bit of information about the system---suffices to constrain its state to within any given two-element subset (e.g. $m_1^+ = \{00, 01\}$ vs. $m_1^- = \{10, 11\}$, as opposed to $m_2^+ = \{00, 10\}$ vs. $m_2^- = \{01, 11\}$). 

\begin{figure}[h] 
 \centering
 \begin{overpic}[width=\textwidth]{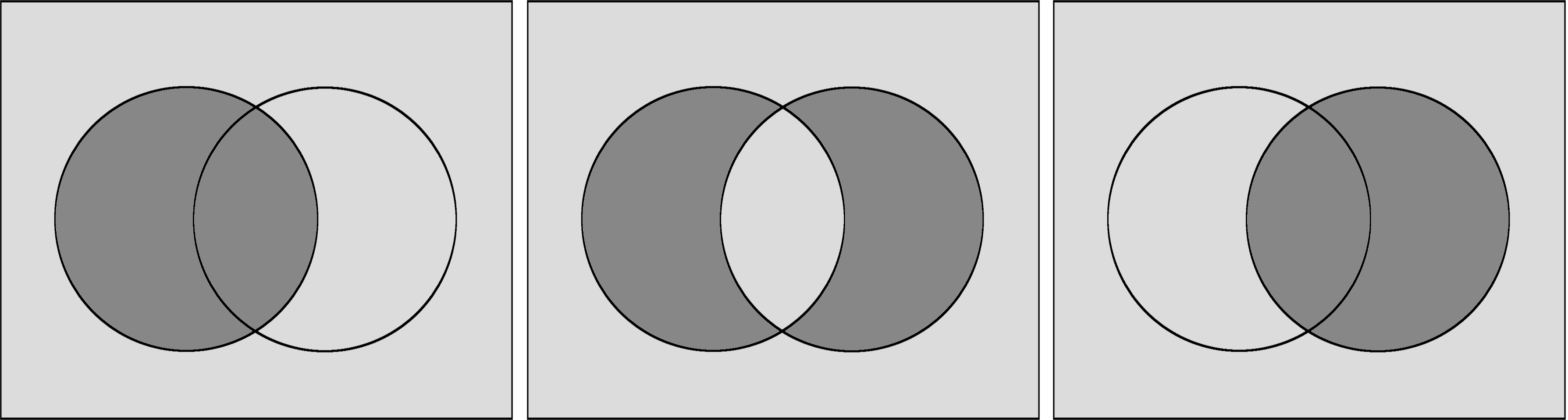}
  \put(1,24){(a)}
  \put(35,24){(b)}
  \put(69,24){(c)}
  \put(6,12){$m_1^+$}
  \put(40,12){$m_{12}^+$}
  \put(91,12){$m_2^+$}
  \put(25,21){$m_1^-$}
  \put(59,21){$m_{12}^-$}
  \put(70,21){$m_2^-$}
\end{overpic}
\caption{Property-calculus in phase space. An elementary system has three properties, $m_1$, $m_2$, and $m_{12} = m_1 \oplus m_2$, only one of which can be definite at any given time. If the system's state lies in the shaded subset of its state space, it has the indicated property, otherwise, it possesses its complement.}
\label{pic:log}
\end{figure}

But this does not exhaust the possibilities---with one bit of knowledge, we may also have information about the relative value of each of the two properties: whether they are the same or opposite, corresponding to the subsets $m_{12}^+ = \{00, 11\}$ or $m_{12}^- = \{10, 01\}$---see Fig.~\ref{pic:log}.

Consequently, an elementary system in this setting possesses three possible properties, only one of which can be definite at any given point. Appreciating the analogy to the qubit, we introduce the notation $m_1 = x_\mathcal{S}$, $m_{12} = y_\mathcal{S}$, $m_2 = z_\mathcal{S}$.

\subsection{Composite Systems}\label{subsec:comp}

New phenomena emerge once we consider composites of elementary systems. A system $\mathcal{A}\otimes\mathcal{B}$, composed of elementary systems $\mathcal{A}$ and $\mathcal{B}$, might be described by definite values for a single property of each of its constituents, e.g. $(x_\mathcal{A},z_\mathcal{B})$. But as stipulated, we can logically combine individual properties. Thus, consider the property $x_\mathcal{AB} = x_\mathcal{A}\oplus x_\mathcal{B}$. The system has the property $x_\mathcal{AB}^+$ if $x_\mathcal{A} = -x_\mathcal{B}$, and $x_\mathcal{AB}^-$ if $x_\mathcal{A} = x_\mathcal{B}$: thus, the property tells us something about the correlation between the $x$-properties of both constituent systems. A state of the form $(x_\mathcal{A}^+,x_\mathcal{AB}^+)$ then tells us that the $x$-value of $\mathcal{A}$ is $1$, and the $x$-value of $\mathcal{B}$ is its opposite, i.e. $-1$.

So far, this might be a perfectly classical situation. But consider now the state $(x_\mathcal{AB}^+,z_\mathcal{AB}^+)$. This refers to a situation in which $x$- and $z$-values (and, by extension, $y$-values) are oppositely aligned; but, the two bits of information available about the system exhausted in this description, no individual $x$- or $z$-value is definite. 

This captures, in the present framework, the phenomenon of \emph{entanglement} (for a related discussion, see \cite{zeilinger1999foundational}). Suppose a measurement of the $x$-value for subsystem $\mathcal{A}$ is performed, and yields the value $+1$: then, we can immediately infer that $x_\mathcal{B}$ must have the value $-1$, due to the correlation. We might then write the state, after measurement on $\mathcal{A}$, as $(x_\mathcal{A}^+,x_\mathcal{AB}^+)$: not only does this measurement tell us $\mathcal{A}$'s value and allow us to predict $\mathcal{B}$'s, but also, all information about the correlation between the $z$-values (and the $y$-values by extension) is destroyed. 

However, as discussed in more depth in \cite{szangolies2020epi}, we must be careful not to simply conflate $(x_\mathcal{A}^+,x_\mathcal{AB}^+)$ and $(x_\mathcal{A}^+,x_\mathcal{B}^-)$: while both states are equivalent in their experimental predictions, they do not support the same counterfactuals. In the state $(x_\mathcal{A}^+,x_\mathcal{AB}^+)$, $\mathcal{B}$'s $x$-value is only defined \emph{conditionally} on $\mathcal{A}$'s value. Thus, if we investigate counterfactual situations---reasoning about what \emph{would have} been the case, \emph{had} $\mathcal{A}$ made a different measurement---a state such as $(x_\mathcal{A}^+,x_\mathcal{B}^-)$ allows us to consider $\mathcal{B}$'s $x$-value independently from $\mathcal{A}$'s, while a state such as $(x_\mathcal{A}^+,x_\mathcal{AB}^+)$ does not. 

To emphasize this distinction, we introduce the notion of a \emph{conditional property} or \emph{conditional event}: the notation $x_\mathcal{B}^-|x_\mathcal{A}^+$ denotes that the value of $\mathcal{B}$'s $x$-observable is $-1$ \emph{given that} $\mathcal{A}$'s $x$-value is $+1$. Hence, we will write the state $(x_\mathcal{A}^+,x_\mathcal{AB}^+)$ equivalently as $(x_\mathcal{A}^+,x_\mathcal{B}^-|x_\mathcal{A}^+)$.

This distinction has important consequences for the discussion of the EPR-argument and Bell's theorem \cite{szangolies2020epi}, calling into question claims that those suffice to establish the non-locality proper of QM (as opposed to the failure of local realism).

Note that the notion of `conditional event' does not necessarily invite any problematic reliance on the notion of observer---not any more, at least, than the relativity of simultaneity in special relativity does: a definite order of events can, in general, only be provided upon specifying an appropriate reference frame relative to which this is the proper order of events. Likewise, with conditional events, a certain value for a given quantity can only be given upon specifying a certain `epistemic frame', that is, a consistent collection of information about a system---whether that information is known to any observer, or not.
\end{appendix}

\bibliography{sn-bibliography}

\end{document}